\documentclass[aip,jmp,reprint,onecolumn]{revtex4-1}

\usepackage{amsmath,empheq,braket,amsthm,amssymb,graphicx,amsfonts,array,xcolor,subfigure,mathrsfs}

\usepackage[utf8]{inputenc}
\usepackage[T1]{fontenc}
\usepackage{mathptmx}
\usepackage{etoolbox}
\usepackage{bm}
\usepackage{dsfont} 
\usepackage{dcolumn}
\usepackage{bbold}
\usepackage[normalem]{ulem}
\usepackage[unicode=true,pdfusetitle, bookmarks=true,bookmarksnumbered=false,bookmarksopen=false,breaklinks=false,pdfborder={0 0 0},backref=false,colorlinks=true,colorlinks=true,linkcolor=blue,citecolor=blue,urlcolor=blue]{hyperref}

\usepackage{soul}

\newtheorem{theo}{Theorem}
\newtheorem{lem}{Lemma}
\newtheorem{cor}{Corollary}

\theoremstyle{definition}

\theoremstyle{remark}
\newtheorem{remark}{Remark}

\newcommand{\tr}[1]{\mathrm{Tr}\left( #1 \right)}

\newcommand{\kb}[2]{\left| #1 \vphantom{#2} \right>\left< #2 \vphantom{#1} \right|} 
\newcommand{\proj}[1]{\kb{#1}{#1}} 

\DeclareMathOperator{\perm}{perm}
\DeclareMathOperator{\Perm}{Perm}
\DeclareMathOperator{\Det}{Det}

\DeclareMathOperator{\sgn}{sgn}
\DeclareMathOperator{\diag}{diag}

\newcommand{\ie}{i.e.}
\newcommand{\eg}{e.g.}

\newcommand\redsout{\bgroup\markoverwith{\textcolor{orange}{\rule[0.5ex]{2pt}{0.4pt}}}\ULon}

\makeatletter
\def\@email#1#2{%
 \endgroup
 \patchcmd{\titleblock@produce}
  {\frontmatter@RRAPformat}
  {\frontmatter@RRAPformat{\produce@RRAP{*#1\href{mailto:#2}{#2}}}\frontmatter@RRAPformat}
  {}{}
}%
\makeatother

\begin{document}

\title{Complementarity between bosonic and fermionic many-body  interferences with partially distinguishable particles}

\author{Marco Robbio}
\affiliation{Centre for Quantum Information and Communication, \'Ecole polytechnique de Bruxelles, CP 165/59, Universit\'e libre de Bruxelles, 1050 Brussels, Belgium}
\affiliation{International Iberian Nanotechnology Laboratory (INL), Av. Mestre Jos\'e Veiga, 4715-330 Braga, Portugal}
\author{Michael~G. Jabbour}
\affiliation{SAMOVAR, T\'el\'ecom SudParis, Institut Polytechnique de Paris, 91120 Palaiseau, France}
\affiliation{Centre for Quantum Information and Communication, \'Ecole polytechnique de Bruxelles, CP 165/59, Universit\'e libre de Bruxelles, 1050 Brussels, Belgium}
\author{Nicolas~J. Cerf}
\affiliation{Centre for Quantum Information and Communication, \'Ecole polytechnique de Bruxelles, CP 165/59, Universit\'e libre de Bruxelles, 1050 Brussels, Belgium}


\begin{abstract}
    It is well known that bosons and fermions exhibit opposite behaviors when experiencing interference, in the sense that bosons have a tendency to bunch whereas fermions have a tendency to antibunch. Recently, this complementarity was mathematically characterized in \href{https://arxiv.org/abs/2312.17709}{[arXiv:2312.17709]}
    by means of an identity relating the transition probabilities of both types of particles in a linear interferometer. Here, we show that such a complementarity still holds even when particles become partially distinguishable, for example, when they have slightly different polarizations or time delays. Namely, we establish a relation that combines bosonic and fermionic multiparticle interferences in an arbitrary linear interferometer, in the presence of partial distinguishability. Incidentally, this also provides a new mathematical identity relating the permanent and determinant of tensors of order 3.
    Importantly, this complementarity has direct operational consequences in quantum metrology.
    Indeed, we show that the correlation matrices for bosonic and fermionic particle number distributions at the output of the interferometer obey a simple sum rule: their sum equals twice the correlation matrix for classical particles.
    This, in turn, constraints the achievable quantum Fisher information in phase-estimation protocols,  highlighting a trade-off whereby greater indistinguishability enhances bosonic sensitivity whereas reduced indistinguishability can benefit fermionic schemes.
\end{abstract}

\maketitle

\section{Introduction}

Unlike classical particles, bosons and fermions exhibit fundamentally different quantum interference behaviors. Bosons are characterized by symmetric wavefunctions and tend to cluster together, or bunch~\cite{Einstein}. In contrast, as a consequence of the Pauli principle~\cite{Pauli}, fermions possess antisymmetric wavefunctions and actively avoid one another, or antibunch.
This intrinsic dichotomy is mathematically evident in their probabilities of transition through linear interferometers. While transition probabilities for both types of particles are calculated using polynomial functions of matrix entries, bosonic probabilities rely on permanents (reflecting their symmetry), whereas fermionic probabilities rely on determinants (reflecting their anti-symmetry). In recent years, there has been an effort in better understanding and characterizing these mathematical objects in the context of quantum interferometry~\cite{Tichy2012a,Tichy2012b,Tichy2017,Jabbour2020,Jabbour2021,jabbour2023bosonfermion}. Indeed, they play a central role in demonstrating a so-called quantum advantage in the context of experiments such as the boson sampling paradigm~\cite{bosonsampling}, which investigates the computational complexity of simulating the scattering of many identical bosons through a multimode linear interferometer.
\par

The seminal work by Hong, Ou, and Mandel~\cite{HOM} on two-photon interference has demonstrated the stark contrast between bosonic bunching and fermionic antibunching, laying the foundation for modern quantum optics research.
Indeed, quantum interference plays a key role in various state-of-the-art photonics technologies, such as quantum computing~\cite{Ladd2010}, quantum cryptography~\cite{Gisin2002}, and
superconducting quantum interference devices~\cite{Vasyukov2013}.
In contemporary experiments, particles often carry internal degrees of freedom, such as polarization or spatiotemporal modes, leading to partial distinguishability~\cite{Tichy_2015}. That is, particles may not be fully indistinguishable in practice as a consequence, for instance, of their occupying slightly different polarization or spatiotemporal modes.
This partial distinguishability introduces deviations from ideal quantum behavior, producing observable effects. 
Mathematically, partial distinguishability is treated explicitly by encoding internal-state overlaps in a Gram matrix, or distinguishability matrix~\cite{Tichy_2015}, which is incorporated into the definition of the particles' transition probabilities [see Eq.~\eqref{3-tensor}]. This mathematical description interpolates continuously between the fully indistinguishable (purely quantum) and fully distinguishable (classical) limits. Working in this framework makes it possible to compare bosonic and fermionic scattering on equal footing: the same interferometric process and the same Gram matrix accomodate both types of statistics. 
\par

Building on this unified framework, we seek to formalize the intuitive notion that bosonic and fermionic many-particle interference are ultimately two faces of the same coin. While these two species are typically analyzed independently due to the divergent computational complexities of permanents and determinants, establishing a direct mathematical link between them is essential for a complete and rigorous understanding of multiparticle quantum statistics. In this work, we demonstrate that the scattering distributions of partially distinguishable bosons and fermions are bound by exact algebraic complementarity relations, directly intertwining permanent- and determinant-type transition probabilities. By analyzing both thermal and fixed-particle (Fock) input states impinging into a linear interferometer, we establish multilinear algebra identities that mathematically formalize this fundamental dichotomy. Beyond these foundational theoretical results, we expose their direct operational consequences for quantum metrology. Specifically, we show that, for any linear interferometer, the correlation matrices for bosonic and fermionic particle number distributions at the output of the interferometer obey a simple sum rule: their sum equals twice the correlation matrix for classical particles.
This relationship places quantitative constraints on the achievable quantum Fisher information in phase-estimation protocols, formally capturing an operational trade-off wherein increased indistinguishability enhances bosonic sensitivity, while the opposite trend appears in fermionic schemes.
\par

The rest of the paper is organized as follows. We begin with a preliminary section (Sec. \ref{Sec:preliminaries}) where we introduce: (i) bosonic and fermionic systems; (ii)  partial distinguishability; (iii) the way probabilities of transition through linear interferometers can be quantified using the familiar determinant and permanent; (iv) and finally some mathematical notions required to prove our results. We then turn to our main results in Section~\ref{sec:main}. Lemma~\ref{th: thermal GF} can be viewed as a new generalization of MacMahon's master theorem~\cite{MacMahon}. Corollary~\ref{cor:Muir_new} gives us a new interpretation of a result of Muir dating from the 19th century~\cite{Muir_1899}, in terms of bosonic and fermionic transition probabilities. Theorem~\ref{cor: Boson/Fermion complementarity} constitutes our most important result: it is a novel fundamental relation that combines bosonic and fermionic multiparticle
interferences in an arbitrary linear interferometer, taking into account partial distinguishability. We then interpret our results in Section~\ref{sec:interpret} in the context of quantum metrology, before proving them in Section~\ref{sec:proof}. Finally, we finish with some conclusions and discussions in Section~\ref{sec:conc}.

\section{Preliminaries}
\label{Sec:preliminaries}

\textit{Bosonic and Fermionic systems-}
An \(m\)-mode bosonic system is represented by the tensor product of \(m\) infinite-dimensional Hilbert spaces, each associated with a pair of bosonic field operators \(\hat{a}_i\) and \(\hat{a}_i^{\dagger}\), where \(i = 1, \ldots, m\) denotes the mode index. These field operators satisfy the canonical bosonic commutation relations:
\begin{equation}
    [\hat{a}_i, \hat{a}_j^\dagger] = \delta_{ij} \mathds{1}, \quad [\hat{a}_i, \hat{a}_j] = 0, \quad  [\hat{a}_i^\dagger, \hat{a}_j^\dagger] = 0,
\end{equation}
where \(\delta_{ij}\) is the Kronecker delta and \(\mathds{1}\) denotes the identity operator in the infinite-dimensional Hilbert space.
Conversely, an \(m\)-mode fermionic system is described by the tensor product of \(m\) two-dimensional Hilbert spaces. Each mode in this system is characterized by a pair of fermionic field operators \(\hat{b}_i\) and \(\hat{b}_i^{\dagger}\), which satisfy the fermionic anti-commutation relations:
\begin{equation}
    \{\hat{b}_i, \hat{b}_j^\dagger\} = \delta_{ij} \mathds{1}, \quad \{\hat{b}_i, \hat{b}_j\} = 0, \quad  \{\hat{b}_i^\dagger, \hat{b}_j^\dagger\} = 0,
\end{equation}
where $\mathds{1}$ is now the identity operator in dimension $2$ (we use the same notation as in the infinite-dimensional case for conciseness).
The quantum states of both bosonic and fermionic systems can be described within their respective Fock spaces. Fock states are denoted as $|\boldsymbol{i}\rangle \coloneqq |i_{1},...,i_{n}\rangle$ for \(\boldsymbol{i} \in \mathbb{N}^m\), where \(\mathbb{N}\) represents the set of natural numbers, including zero.

Given an $m$-mode linear interferometer, its effect on the mode operators is characterized in phase space by a unitary matrix $U$ of dimension $m$,
which acts on the vectors of field operators $\boldsymbol{\hat{c}} \coloneqq \left(\hat{c}_1, \cdots, \hat{c}_m \right)^T$ as follows:
\begin{equation}\label{eq:inter}
    \boldsymbol{\hat{c}} \rightarrow U^{\dagger} \boldsymbol{\hat{c}} =
    \left( \sum_{j=1}^m U^{\dagger}_{1,j} \hat{c}_j, \sum_{j=1}^m U^{\dagger}_{2,j} \hat{c}_j, \hdots, \sum_{j=1}^m U^{\dagger}_{m,j} \hat{c}_j \right)^T,
\end{equation}
where $\hat{c}_i = \hat{a}_i$ (or $ \hat{b}_i$) for a bosonic (or fermionic) mode.
%
Equivalently, the effect of the linear interferometer can be characterized in state space by a unitary matrix $\hat{U}$ via the relation:
\begin{equation} \label{eq:Uc}
    \hat{U}^{\dagger} \hat{c}_{i} \hat{U} = \sum_{j} U^{\dagger}_{i,j} \hat{c}_{j}, \quad \forall i,
\end{equation}
that is, $\hat{U}^{\dagger} \, \boldsymbol{\hat{c}} \, \hat{U}=U^{\dagger} \boldsymbol{\hat{c}}$ in a concise vectorial form (with a slight abuse of notation).
Note that the structure of the matrix $\hat{U}$ depends on whether we are dealing with bosons or fermions. Indeed, in the former case, it acts on the tensor product of $m$ infinite-dimensional Hilbert spaces, while in the later case, it acts on the tensor product of $m$ two-dimensional Hilbert space (notice thus the slight abuse of notation on the left-hand side of Eq.~\eqref{eq:Uc}, where we should write a tensor product of $\hat{c}_{i}$ and identities). We choose to use the same notation for $\hat{U}$ for both types of particles for conciseness, but its nature should be clear from the context. Note however that the matrix $U$ is the same for both types of particles, which is important in what follows.

\medskip

\textit{Partial distinguishability--}
To introduce the concept of partially distinguishable particles, we begin with the simplest case, which describes two particles entering two different input arms of a beam splitter (we will use a bosonic system for the example, but the same can be said for a fermionic one). The initial state can be expressed as~\cite{GSO}:
\begin{equation}\label{2-photon-input}
|\Psi_{\text{in}}\rangle=\hat{a}_{1,\phi_{1}}^{\dagger}\hat{a}_{2,\phi_{2}}^{\dagger}|vac\rangle,
\end{equation}
where $\ket{vac}$ represents a tensor product of vacuum states (the ground state of the free Hamiltonian), while the operator $\hat{a}_{i,\phi_{j}}^{\dagger}$ generates a particle in spatial mode $i$ with internal degree of freedom (\eg, frequency, polarization, or spin) described by a wavefunction $|\phi_{j}\rangle$. This assumption is coherent with the common scenario in photonic where it is usually assumed that each photon has a high spectral purity and a specific polarization, meaning that its internal degrees of freedom can be accurately described by pure states.
While the states $\ket{\phi_1}$ and  $\ket{\phi_2}$ are in general non-orthogonal, we can express them in a two-dimensional orthonormal basis $\{\ket{u}\}$ as 
\begin{align}
|\phi_{1}\rangle=\sum_{u=1}^{2}c_{1,u}|u\rangle, \quad
|\phi_{2}\rangle=\sum_{u=1}^{2}c_{2,u}|u\rangle,
\end{align}
where the $c_{i,u}$ are complex numbers.
Consequently, we can express the initial state as:
\begin{equation}\label{eq:partialdis}
|\Psi_{\text{in}}\rangle={\bigg (}\sum_{u,v}c_{1,u}\, c_{2,v}\, \hat{a}^{\dagger}_{1,u}\, \hat{a}^{\dagger}_{2,v}{\bigg )}|vac\rangle, 
\end{equation}
where the operators $\hat{a}_{i,u}$ obey the usual commutations relations $[\hat{a}_{i,u}, \hat{a}_{j,v}^{\dagger}]= \delta_{i,j}\delta_{u,v}$ for bosons and anti-commutation relations $\{\hat{b}_{i,u}, \hat{b}_{j,v}^{\dagger}\} = \delta_{i,j}\delta_{u,v}$ for fermions. It is well known that if this state goes through a 50-50 beam-splitter which couples only the spatial modes, the probability of observing one particle in each output mode (\ie, the antibunching probability) depends on the overlap $|\braket{\phi_1|\phi_2}|^2$. 

More generally, the linear interference of multiple partially distinguishable photons is usually described by a unitary which acts on creation operators in an analogous way as in Eq.~\eqref{eq:Uc}, \ie~as 
\begin{equation} \label{eq:qaddn}
    \hat{U}^{\dagger} \hat{a}^{\dagger}_{i, u} \hat{U}= \sum_{j} U_{i,j}^{\dagger} \hat{a}^{\dagger}_{j, u},  
\end{equation}
coupling the different spatial modes $i$ and $j$, but leaving internal degrees of freedom $u$ unchanged (the dimensions of $U$ and $\hat{U}$ now depend on both the number of spatial modes and internal degrees of freedom). Assuming detectors can only count the total number of particles in a given spatial mode (independently of their internal degrees of freedom), the particles-counting statistics of such an experiment depends only on the unitary $U$ and the distinguishability matrix $S$, whose elements are defined as follows:
\begin{equation}\label{eq:S}
    S_{i,j}=\langle \phi_{i}|\phi_{j}\rangle.
\end{equation}
The above matrix is the Gram matrix associated with the pairwise overlaps of the internal states $\ket{\phi_j}$~\cite{Tichy2015}.  
When $S=I$, where $I$ will be the identity matrix of dimension $m$ throughout, we recover the case of fully distinguishable particles, which occurs in the classical case. In contrast, the case of fully indistinguishable particles is described by a Gram matrix whose elements satisfy $S_{i,j}= 1$ for all $i$ and $j$.
Throughout the discussion, we assume that each spatial mode \( i \) carries an associated internal degree of freedom \( \phi_i \). This assumption is close to the photonic scenario in which different sources are used to generate different states, and each source carries its own internal degree of freedom. Consequently, the dimension of the Gram matrix scales with the size of the interferometer. This approach, although common in the literature, is not fully general; in fact, one could consider states of the type
\begin{equation}
    |\psi\rangle=\frac{1}{\sqrt{1+|\langle \phi_{1}|\phi_{2}\rangle|^{2}}} \hat{a}_{1,\phi_{1}}^{\dagger} \hat{a}_{1,\phi_{2}}^{\dagger}|vac\rangle,
\end{equation}
where two photons which are partially distinguishable occupy the same mode, for which our assumption does not hold\cite{steinmetz2024simulatingimperfectquantumoptical}. 
Although the specific form of the state that describes the internal degrees of freedom is not crucial for our discussion, it is important to note that with $m$ modes, we can always choose an orthonormal basis of dimension at least $d\leq m$  to describe all internal states $\ket{\phi_i}$, where $d=\text{rank}(S)$ is the dimension of the Hilbert space spanned by the states $\{\ket{\phi_i}\}$.
We point the interested reader to Ref.~\onlinecite{Robbio2024} for more details about the mathematics of partial distinguishability for bosons.

\medskip

\textit{Transition probabilities and their relations to permanents and determinants--}
We define the bosonic transition probabilities for a linear passive interferometer described by the matrix $U$ as:
\begin{equation}\label{eq:transitionB}
    B_{\boldsymbol{k}\to \boldsymbol{i}}=|\langle \boldsymbol{k}|\hat{U}|\boldsymbol{i}\rangle|^{2}.
\end{equation}
This is equivalent to compute the probability that given an input occupation vector $\boldsymbol{k}=(k_{1},...,k_{m})\in \mathbb{N}^{m}$ which evolve under the effect of the interferometer described by the matrix $U$, we measure a mode occupation vector $\boldsymbol{i}=(i_{1},...,i_{m})\in \mathbb{N}^{m}$. The output measurement is performed without considering the internal degrees of freedom, meaning we are measuring the operator $\hat{n}_{i_{j}}=\sum_{u}\hat{a}_{i_{j},u}^{\dagger} \hat{a}_{i_{j},u}$. This is the common approach used in state of the art experiments, where detectors are usually insensitive to internal degrees of freedom as polarization, frequency and time delays. With the same formalism we define the fermionic transition probabilities as
\begin{equation}\label{eq:transitionF}
    F_{\boldsymbol{k}\to \boldsymbol{i}}=|\langle \boldsymbol{k}|\hat{U}|\boldsymbol{i}\rangle|^{2},
\end{equation}
where the same applies but we are measuring the operator $\sum_{u}\hat{b}_{i_{j},u}^{\dagger}\hat{b}_{i_{j},u}$ (also remember that $\hat{U}$ is different whether we are dealing with bosons or fermions, so that the transition probabilities of Eqs.~\eqref{eq:transitionB} and~\eqref{eq:transitionF} are indeed different).  Note that, under the assumption that the $i$\textsuperscript{th} mode carries an internal degree of freedom $\phi_i$, fermionic states with more than one fermion in the same input mode are forbidden by the Pauli exclusion principle. However, output configurations with more than one particle in the same mode are allowed, since fermions occupying the same output mode but having different internal degrees of freedom are not subject to the Pauli constraint.

One can show that both bosons and fermions transition probabilities can be rewritten in terms of permanents and determinants, whose respective definitions we recall here:
\begin{equation}
\begin{split}
    \perm(A)=\sum_{\sigma\in \mathcal{S}_n}\prod_{j}A_{\sigma(j),j}, \quad
    \det(A)=\sum_{\sigma\in \mathcal{S}_n}\prod_{j}\sgn(\sigma)A_{\sigma(j),j},
\end{split}
\end{equation}
where we indicates the permanent of the matrix $A\in \mathbb{C}^{n\times n}$ with the symbol $\perm(A)$ and the determinant with $\det(A)$, and where $\mathcal{S}_n$ denotes the symmetric group. In the case of indistinguishable particles, given the unitary matrix $U$, we define the scattering matrix $U^{(\boldsymbol{m},\boldsymbol{j})}$ to be the submatrix of $U$ with rows (columns) characterized by the vector $\boldsymbol{m}$ ($\boldsymbol{j})$. Since unoccupied input and output modes do not affect particle scattering, we define the effective scattering matrix $U^{({\boldsymbol{m},\boldsymbol{j}})}$ as the relevant submatrix of $U$ that contains those rows and columns corresponding to the initially and finally populated modes, so that the multiplicity of rows and columns reflects the respective population~\cite{Tichy2015}. The transition probabilities can be then rewritten as
\begin{equation}\label{eq:BperFdet_indis}
    B_{\boldsymbol{m}\to \boldsymbol{j}}=\frac{1}{\boldsymbol{m}!\boldsymbol{j}!}|\perm(U^{({\boldsymbol{m},\boldsymbol{j}})})|^{2}, \quad
    F_{\boldsymbol{m}\to \boldsymbol{j}}=\frac{1}{\boldsymbol{m}!\boldsymbol{j}!}|\det(U^{({\boldsymbol{m},\boldsymbol{j}})})|^{2}.
\end{equation}
When dealing with partial distinguishable particles, we need to modify the above definitions accordingly, in particular by including the Gram matrix $S$ of Eq.~\eqref{eq:S} in the story. The most compact way to do so is by introducing the following $n^{3}-$dimensional $3$-tensor:
\begin{equation}\label{3-tensor}
    W_{k,l,t}^{({\boldsymbol{m},\boldsymbol{j}})}=U_{k,t}^{({\boldsymbol{m},\boldsymbol{j}})}(U_{l,t}^{({\boldsymbol{m},\boldsymbol{j}})})^{*}S_{l,k}.
\end{equation}
Now, the permanent and determinant of the $3$-tensor $W$ are defined as follows:
\begin{equation}
    \Perm(W)=\sum_{\sigma_{1},\sigma_{2}\in \mathcal{S}_n}\prod_{j}W_{\sigma_{1}(j),\sigma_{2}(j),j}, \quad \Det(W)=\sum_{\sigma_{1},\sigma_{2}\in \mathcal{S}_n}\prod_{j}\sgn(\sigma_{1})\sgn(\sigma_{2})W_{\sigma_{1}(j),\sigma_{2}(j),j},
\end{equation}
where we have used a capital letter to distinguish it from its matrix version.
If particles are partially distinguishable, it has been shown in Ref.~\onlinecite{Tichy_2015} that the transition probabilities now depend on the Gram matrix $S$ through the following modified versions of Eq.~\eqref{eq:BperFdet_indis}:
\begin{equation}\label{eq:transitions}
    B_{\boldsymbol{m}\to \boldsymbol{j}}
    = \frac{1}{\boldsymbol{m}!\boldsymbol{j}!}\Perm(W^{({\boldsymbol{m},\boldsymbol{j}})}), \quad F_{\boldsymbol{m}\to \boldsymbol{j}}
    = \frac{1}{\boldsymbol{m}!\boldsymbol{j}!}\Det(W^{({\boldsymbol{m},\boldsymbol{j}})}).
\end{equation}

Notice that the $3$-tensor $W$ reduces to the more commonly known forms in the case of fully indistinguishable and fully distinguishable particles. In particular, we have:
\begin{equation}\label{eq: BS transitions}
    B_{\boldsymbol{m}\to \boldsymbol{j}}=\frac{1}{\boldsymbol{m}!\boldsymbol{j}!}\begin{cases} \perm(|U^{({\boldsymbol{m},\boldsymbol{j}})}|^{2}), & \text{for distinguishable bosons,} \\
    |\perm(U^{({\boldsymbol{m},\boldsymbol{j}})})|^{2}, & \text{for indistinguishable bosons,}
    \end{cases}
\end{equation}
\begin{equation}\label{eq: FS transitions}
    F_{\boldsymbol{m}\to \boldsymbol{j}}=\frac{1}{\boldsymbol{m}!\boldsymbol{j}!}\begin{cases} \perm(|U^{({\boldsymbol{m},\boldsymbol{j}})}|^{2}), & \text{for distinguishable fermions,} \\
    |\det(U^{({\boldsymbol{m},\boldsymbol{j}})})|^{2}, & \text{for indistinguishable fermions,}
    \end{cases}
\end{equation}
where we define the elements of the matrix $|A|^{2}$ as $(|A|^{2})_{i,j}=|A_{i,j}|^{2}$, which is different from the usual modulus square of a matrix. 

Note that, in the case of fully distinguishable particles, the definitions of the transition probabilities coincide.
This is due to the fact that quantum interferences are effectively absent in this case, as it is only present if there is an overlap between the internal degrees of freedom. As a consequence, particles become effectively classical in that case, and their nature does not matter anymore. We will refer to such a probability distribution as $P_{\boldsymbol{k}\to \boldsymbol{i}}$. Nevertheless, in our model, we assume that two particles in the same input mode are indistinguishable, resulting in a straightforward difference: while a mode can be occupied by an arbitrary number of indistinguishable bosons, this does not happen for fermions, due to the Pauli exclusion principle. For additional details, see Appendix~\ref{sec: classical fermions and bosons}. Mathematically, we can explain the fact that both the fermionic and bosonic distributions are described by $\perm(|U|^{2}),$ for $S_{ij}=\delta_{ij}$, by simply noticing that in this case
\begin{align}
    & \Det(W^{(\boldsymbol{m},\boldsymbol{j})}){\bigg |}_{S=I} =\sum_{\sigma_{1},\sigma_{2}\in \mathcal{S}_n}\prod_{j}\sgn(\sigma_{1})\sgn(\sigma_{2})W_{\sigma_{1}(j),\sigma_{2}(j),j}^{(\boldsymbol{m},\boldsymbol{j})}\delta_{\sigma_{1}(j),\sigma_{2}(j)}
    = \sum_{\sigma_{1}\in \mathcal{S}_n}\prod_{j}W_{\sigma_{1}(j),\sigma_{1}(j),j}^{(\boldsymbol{m},\boldsymbol{j})}=\perm(|U^{(\boldsymbol{m},\boldsymbol{j})}|^{2}), \\
    & \Perm(W^{(\boldsymbol{m},\boldsymbol{j})}){\bigg |}_{S=I} =\sum_{\sigma_{1},\sigma_{2}\in \mathcal{S}_n}\prod_{j}W_{\sigma_{1}(j),\sigma_{2}(j),j}^{(\boldsymbol{m},\boldsymbol{j})}\delta_{\sigma_{1}(j),\sigma_{2}(j)}
    = \sum_{\sigma_{1}\in \mathcal{S}_n}\prod_{j}W_{\sigma_{1}(j),\sigma_{1}(j),j}^{(\boldsymbol{m},\boldsymbol{j})}=\perm(|U^{(\boldsymbol{m},\boldsymbol{j})}|^{2}),
\end{align}
where we write $I$ for the identity matrix of dimension $m$.

The interference scenarios described above are commonly referred to as boson/fermion sampling~\cite{aaronson2010computational} (we will refer to them as BS and FS, respectively). The boson sampling scheme is strongly believed to implement computing tasks which are hard to implement with classical computers by using far fewer physical resources than a full linear-optical quantum computing setup. This advantage makes it an ideal candidate for demonstrating the power of quantum computation in the near term. On the other hand, due to the known efficient algorithms for computing determinants, the fermionic version is less interesting complexity-wise. The addition of partial distinguishability (in particular models) has shown to make the problem more approachable to classical simulation~\cite{hoven2024efficientclassicalalgorithmsimulating}. In the limit of fully distinguishable particles, the problem is classically solvable in polynomial time.

Let us define a different version of this task, which we will refer to as thermal sampling (TBS for bosons and TFS for fermions), where we change the input states to thermal states of the form
\begin{align}
    \rho^{(b)}(\boldsymbol{x})&=\bigotimes_{i=1}^{m}{\bigg (}(1-x_{i})\sum_{k=0}^{\infty}x_{i}^{k}|k\rangle\langle k|{\bigg )}, \\
    \rho^{(f)}(\boldsymbol{x})&=\bigotimes_{i=1}^{m}{\bigg (}\frac{1}{1+x_{i}}|0\rangle\langle 0|+\frac{x_{i}}{1+x_{i}}|1\rangle\langle 1|{\bigg )},
\end{align} 
where the vector $\boldsymbol{x}=(x_{1},...,x_{m}) \in [0,1)^{m}$ characterizes the bosonic and fermionic thermal states.
We introduce the task of thermal sampling for two distinct reasons: (i) We will show that the associated transition probabilities are related to a known mathematical problem which connects permanents and determinants, then leading us to prove novel relations for these quantities; (ii) We can exploit the formalism of TFS/TBS to easily derive novel results for the FS and BS cases.
The transition probabilities in the TBS and TFS setups can respectively be defined as follows:
\begin{equation}\label{eq:thermal_overlap}
    b(\boldsymbol{x},\boldsymbol{j}) = \tr{\hat{U}\rho^{(b)}(\boldsymbol{x})\hat{U}^{\dagger}\proj{\boldsymbol{j}} },
    \quad
    f(\boldsymbol{x},\boldsymbol{j}) = \tr{\hat{U}\rho^{(f)}(\boldsymbol{x})\hat{U}^{\dagger}\proj{\boldsymbol{j}}}.
\end{equation}
Later in the discussion, we will compare bosonic and fermionic thermal states characterized by the same parameter \( \boldsymbol{x} \), since they share a closely related mathematical structure. Nevertheless, one should remember that the average number of particles is a function of $ \boldsymbol{x}$: for bosons, mode \( i \) has average occupation \( \frac{x_i}{1 - x_i} \), whereas for fermions it is \( \frac{x_i}{1 + x_i} \).
It can be shown~\cite{chakhmakhchyan2017quantum} that the above transition probabilities can be rewritten in terms of permanent and determinant as follows:
\begin{equation}\label{eq: transition prob-thermal}
    b(\boldsymbol{x},\boldsymbol{j})=\Bigg(\prod_{i=1}^M (1-x_i)\Bigg) \frac{\perm\!\left(M^{(\boldsymbol{j},\boldsymbol{j})}\right)}{\boldsymbol{j}!}, \quad f(\boldsymbol{x},\boldsymbol{j})=\Bigg(\prod_{i=1}^M \frac{1}{1+x_i}\Bigg)\det\!\left(M^{(\boldsymbol{j
    },\boldsymbol{j})}\right),
\end{equation}
where $M=U\diag(\boldsymbol{x})U^{\dagger}$. This setup is known to give an easy to classically simulate sampling process, both for bosons and fermions.
As we will show later, our main result will allow us in particular to recover a relation between the permanent and determinant that was proven by Muir at the end of the 19th century~\cite{Muir_1899} (see Eq.~\eqref{eq: Muir} below). Interestingly, the relation was never stated in terms of bosonic and fermionic transition probabilities, which is what our main result allows us to do.

\medskip

\textit{Generating functions--}
A powerful tool for the study of multivariate sequences, such as the transition probabilities of Eq.~\eqref{eq:BperFdet_indis}, can be found in \emph{generating functions} (GF) (see for instance Refs.~\onlinecite{Jabbour2021,jabbour2023bosonfermion} for applications in theoretical quantum optics). The GFs of the probabilities are defined as follows:
\begin{align}\label{eq: definition of GF}
    G_{B}(\boldsymbol{x},\boldsymbol{y})=\sum_{\boldsymbol{m},\boldsymbol{j}}B_{\boldsymbol{m}\to\boldsymbol{j}}, \boldsymbol{x}^{\boldsymbol{m}}\boldsymbol{y}^{\boldsymbol{j}}, \quad 
    G_{F}(\boldsymbol{x},\boldsymbol{y})=\sum_{\boldsymbol{m},\boldsymbol{j}}F_{\boldsymbol{m}\to\boldsymbol{j}}, \boldsymbol{x}^{\boldsymbol{m}}\boldsymbol{y}^{\boldsymbol{j}},
\end{align}
with the use of the notation $\boldsymbol{x}^{\boldsymbol{m}}=\prod_{i}x_{i}^{m_{i}}$. We mention at this point that the transition probabilities can be recovered from the GFs through the following relations:
\begin{align} \label{eq:GftoProba_bos}
    B_{\boldsymbol{m}\to \boldsymbol{j}}=\frac{1}{\boldsymbol{m}!\boldsymbol{j}!}\frac{\partial^{\boldsymbol{m}}\partial^{\boldsymbol{j}}}{\partial \boldsymbol{x}^{\boldsymbol{m}}\partial\boldsymbol{y}^{\boldsymbol{j}}}G_{B}(\boldsymbol{x},\boldsymbol{y}){\bigg |}_{x_{i}=0,y_{i}=0, \forall i},
    \quad
    F_{\boldsymbol{m}\to \boldsymbol{j}}=\frac{1}{\boldsymbol{m}!\boldsymbol{j}!}\frac{\partial^{\boldsymbol{m}}\partial^{\boldsymbol{j}}}{\partial \boldsymbol{x}^{\boldsymbol{m}}\partial\boldsymbol{y}^{\boldsymbol{j}}}G_{F}(\boldsymbol{x},\boldsymbol{y}){\bigg |}_{x_{i}=0,y_{i}=0, \forall i},
\end{align}
where we define $\frac{\partial^{\boldsymbol{m}}}{\partial \boldsymbol{x}^{\boldsymbol{m}}}=\prod_{i}\frac{\partial^{m_{i}}}{\partial x_{i}^{m_{i}}}$ and $\boldsymbol{m}!=\prod_{i}m_{i}!$.
Notice that, although mathematically acceptable, some mode occupations are not physical for fermions due to the Pauli exclusion principle, which indeed leads to a zero probability from properties of the determinant.

In the same way, we can approach the TBS problem (and its fermionic counterpart) by means of GFs, where this time we assume to have $m$ input Gaussian thermal states, with average particle number related to $\boldsymbol{x}$. 
We define the two generating functions in the TBS and TFS as follows:
\begin{equation} \label{eq:gBF}
    g_{B}(\boldsymbol{x},\boldsymbol{y}) = \sum_{\boldsymbol{j}}b(\boldsymbol{x},\boldsymbol{j}) \, \boldsymbol{y}^{\boldsymbol{j}},
    \quad
    g_{F}(\boldsymbol{x},\boldsymbol{y}) = \sum_{\boldsymbol{j}}f(\boldsymbol{x},\boldsymbol{j}) \, \boldsymbol{y}^{\boldsymbol{j}}.
\end{equation}
Notice that, in contrast with the previous case (BS/FS), to retrieve the probabilities, we perform derivatives only with respect to the $y_{i}$ variables
\begin{align} \label{eq:GftoProba_bos_thermal}
    b(\boldsymbol{x},\boldsymbol{j})=\frac{1}{\boldsymbol{j}!}\frac{\partial^{\boldsymbol{j}}}{\partial\boldsymbol{y}^{\boldsymbol{j}}}g_{B}(\boldsymbol{x},\boldsymbol{y}){\bigg |}_{y_{i}=0, \forall i},
    \quad
     f(\boldsymbol{x},\boldsymbol{j})=\frac{1}{\boldsymbol{j}!}\frac{\partial^{\boldsymbol{j}}}{\partial\boldsymbol{y}^{\boldsymbol{j}}}g_{F}(\boldsymbol{x},\boldsymbol{y}){\bigg |}_{y_{i}=0, \forall i},
\end{align}
since the $\boldsymbol{x}$ variables now specify the input thermal states.

\section{Main results: Boson--fermion complementarity\label{sec:main}}

Bosonic and fermionic many-particle interference present two faces of the same physical coin: bosons tend to bunch, fermions antibunch, and classical (fully distinguishable) particles sit between these extremes. The section that follows makes this qualitative dichotomy precise. Once internal-state overlaps—encoded in the Gram (distinguishability) matrix $S$—and the unitary action of an arbitrary linear interferometer $U$ are taken into account, bosonic and fermionic output statistics become linked by exact algebraic identities. These relations formalize a notion of complementarity between the two quantum statistics, connecting the mathematical structures that govern their interference patterns. We now present several results describing bosonic and fermionic interference in the presence of partial distinguishability, but as already mentioned earlier, we leave their proofs to Section~\ref{sec:proof}.

As already hinted at, we will make heavy use of the generating functions in order to prove our main results. We thus begin by showing how the GFs defined in Eq.~\eqref{eq:gBF} can be written in terms of a determinant.
Incidentally, the following Lemma can be understood as a generalization of MacMahon's master theorem~\cite{MacMahon}, now involving partial distinguishability (see also Eq.~(30) in Ref.~\onlinecite{jabbour2023bosonfermion}).
The proof is provided in Section~\ref{sec:proofLemma}.
\begin{lem}\label{th: thermal GF}
    For every unitary $U$ and Gram matrix $S$, we have
    \begin{align}
        g_{F}(\boldsymbol{x},\boldsymbol{y}) &=\frac{\det(I+(U^{\dagger}YU\odot S)X)}{\det(I+X)},\label{eq: thermal Gf}\\
        g_{B}(\boldsymbol{x},\boldsymbol{y}) &=\frac{\det(I-X)}{\det(I-(U^{\dagger}YU\odot S)X)}.\label{eq: thermal Gb}
    \end{align}
\end{lem}

The first application of Lemma~\ref{th: thermal GF} is both a simple proof (for positive semi-definite matrices) and a reinterpretation of a result by Muir from the end of the 19th century~\cite{Muir_1899}. Muir showed the following relation between permanents and determinants:
\begin{equation}\label{eq: Muir}
    \sum_{\boldsymbol{i}}^{\boldsymbol{j}}\, \frac{(-1)^{|\boldsymbol{i}|}}{(\boldsymbol{j}-\boldsymbol{i})!}\,\perm(M^{(\boldsymbol{j}-\boldsymbol{i}),(\boldsymbol{j}-\boldsymbol{i})})\, \det(M^{(\boldsymbol{i},\boldsymbol{i})})=\delta_{\boldsymbol{j},0},
\end{equation}
where we have introduced the notation $\sum_{\boldsymbol{i}}^{\boldsymbol{j}}$ for the summations over all the elements of vector $\boldsymbol{i}\in \mathbb{N}^{m}$ such that $ \boldsymbol{j}\in \mathbb{N}^{m}$ and $\boldsymbol{j}-\boldsymbol{i}$ are element-wise non-negative vectors.
Note that we could have taken the summations over $\boldsymbol{i}\in \{0,1\}^{m}$, since the determinant would be zero in case of repeated rows and columns.
Equivalently, Lemma~\ref{th: thermal GF} easily implies the following rewriting of Eq.~\eqref{eq: Muir}, whose proof we give in Section~\ref{sec:proofCorr}.
\begin{cor}[Thermal complementarity]\label{cor:Muir_new}
    Let $\boldsymbol{x} \in [0,1)^{m}$ be a vector characterizing an input thermal state impinging into an interferometer described by a unitary matrix $U$ in phase space, and let $\boldsymbol{j} \in \mathbb{N}^N$ be a vector of output occupation numbers.
    Let $b(\boldsymbol{x},\boldsymbol{i})$ and $f(\boldsymbol{x},\boldsymbol{i})$ denote the overlaps describing the associated thermal sampling scheme, as described in Eq.~\eqref{eq:thermal_overlap}. Then,
    \begin{equation}\label{eq:Muir2}
        \sum_{\boldsymbol{i}}^{\boldsymbol{j}}(-1)^{|\boldsymbol{i}|} b(\boldsymbol{x},\boldsymbol{j}-\boldsymbol{i}) f(\boldsymbol{x},\boldsymbol{i})
        =\Bigg(\prod_{i=1}^m \frac{1-x_{i}}{1+x_{i}}\Bigg)\delta_{\boldsymbol{j},0}.
    \end{equation}
\end{cor}
\noindent Using Eq.~\eqref{eq: transition prob-thermal}, we see that Eqs.~\eqref{eq: Muir} and~\eqref{eq:Muir2} are in fact equivalent for positive-semidefinite matrices. Interestingly, we now understand how Muir's relation~\eqref{eq: Muir} was ultimately describing the complementarity between bosons and fermions in the TFS/TBS settings.
The physical interpretation of Eq.~\eqref{eq:Muir2} can be understood by comparing it with the case of classical particles. Since classical particles do not interfere, the corresponding transition probabilities can be computed using a convolution (see for instance the discussion in Section~II.D of Ref.~\onlinecite{Jabbour2021} for more details):
\begin{equation}\label{eq: classical convolution}
    P_{\boldsymbol{i}\to \boldsymbol{m}}=\sum_{\boldsymbol{k}}P_{\boldsymbol{j}\to\boldsymbol{k}}P_{\boldsymbol{i}-\boldsymbol{j}\to \boldsymbol{m}-\boldsymbol{k}}, \quad \forall \boldsymbol{j}\leq \boldsymbol{i}.
\end{equation}
If we now consider the setting of TBS and TFS, Eq.~\eqref{eq:Muir2} tell us that, for an output configuration having $\boldsymbol{j}\neq 0$, we have
\begin{equation}
    b(\boldsymbol{x},\boldsymbol{j}) f(\boldsymbol{x},\boldsymbol{0}) = \sum_{\boldsymbol{i}:\boldsymbol{i}\neq 0}^{\boldsymbol{j}} (-1)^{|\boldsymbol{i}|+1} f(\boldsymbol{x},\boldsymbol{i}) b(\boldsymbol{x},\boldsymbol{j}-\boldsymbol{i}),
\end{equation}
or,
\begin{equation}
    b(\boldsymbol{x},\boldsymbol{j})  = \left(\prod_{n} (1+x_{n}) \right) \sum_{\boldsymbol{i}:\boldsymbol{i}\neq 0}^{\boldsymbol{j}} (-1)^{|\boldsymbol{i}|+1} f(\boldsymbol{x},\boldsymbol{i}) b(\boldsymbol{x},\boldsymbol{j}-\boldsymbol{i}).
\end{equation}
We see a clear resemblance with Eq.~\ref{eq: classical convolution}. The main difference is that the quantity $b(\boldsymbol{x},\boldsymbol{j})$ is now given by the convolution of the two distinct objects $b(\boldsymbol{x},\boldsymbol{j})$ and $(-1)^{|\boldsymbol{i}|+1} f(\boldsymbol{x},\boldsymbol{i})$, along with a renormalization factor. This analogy with the classical case will be even more striking in the case of the usual boson and fermion sampling schemes, which we consider next.

The complementarity observed in thermal states establishes a strict constraint between bosonic and fermionic interference. However, because thermal distributions are statistical mixtures encompassing all possible particle numbers, this thermal relationship is fundamentally an average. The fact that an exact cancellation occurs across the entire thermal ensemble strongly implies the existence of a deeper, mechanism at the level of individual multiparticle scattering events. To isolate and understand this underlying mechanism, we turn our focus from thermal ensembles to fixed-particle-number scenarios—specifically, standard Boson and Fermion Sampling (BS and FS). By removing the statistical averaging inherent to thermal states, we can directly analyze the core transition probabilities governing these processes.
We show the following in Section~\ref{sec:proofTheo}.
\begin{theo}[Boson/Fermion complementarity]\label{cor: Boson/Fermion complementarity}
    Let $\boldsymbol{k}, \boldsymbol{i} \in \mathbb{N}^m$ be vectors of input and output occupation numbers and $B_{\boldsymbol{k}\to\boldsymbol{i}}$ (or $F_{\boldsymbol{k}\to\boldsymbol{i}}$) denote the bosonic (or fermionic) transition probability from $\ket{\boldsymbol{k}}$ to $\ket{\boldsymbol{i}}$ via a given linear interferometer characterized by a unitary matrix $U$ in phase-space, in presence of partial distinguishability described by a Gram matrix $S$. Then,
    \begin{equation}\label{eq: Complementarity BS/FS}
        \sum_{\boldsymbol{m},\boldsymbol{j}\in\mathbb{N}^{n}}^{\boldsymbol{k},\boldsymbol{i}}(-1)^{|\boldsymbol{j}|}F_{\boldsymbol{m}\to\boldsymbol{j}}B_{\boldsymbol{k}-\boldsymbol{m}\to\boldsymbol{i}-\boldsymbol{j}}
        = \delta_{\boldsymbol{k},\boldsymbol{0}}\delta_{\boldsymbol{i},\boldsymbol{0}}.
    \end{equation}
    Equivalently, for the $3$-tensor $W$ whose elements are defined as $W_{k,l,t}^{({\boldsymbol{m},\boldsymbol{j}})}=U_{k,t}^{({\boldsymbol{m},\boldsymbol{j}})}(U_{l,t}^{({\boldsymbol{m},\boldsymbol{j}})})^{*}S_{l,k}$, we have
    \begin{equation}\label{eq:DetPer}
        \sum_{\boldsymbol{m},\boldsymbol{j}\in\mathbb{N}^{n}}^{\boldsymbol{k},\boldsymbol{i}}(-1)^{|\boldsymbol{j}|}\frac{\Det(W^{(\boldsymbol{m},\boldsymbol{j})})}{\boldsymbol{m}!\boldsymbol{j}!}\frac{\Perm(W^{(\boldsymbol{k-m},\boldsymbol{i-j})})}{\boldsymbol{(k-m)}!\boldsymbol{(i-j)}!}
        = \delta_{\boldsymbol{k},\boldsymbol{0}}\delta_{\boldsymbol{i},\boldsymbol{0}}.
    \end{equation}
\end{theo}

If we compare Eq.~\eqref{eq: Complementarity BS/FS} with Eq.~\eqref{eq:Muir2}, the first thing to note is that the expression is now given by a double convolution rather than a single one. In Eq.~\eqref{eq:Muir2}, the convolution is over indices $\boldsymbol{i}$ characterizing outputs only, while the real number $\boldsymbol{x}$ characterizing the thermal input is fixed. In contrast, in Eq.~\eqref{eq: Complementarity BS/FS}, the convolution is now over two indices $\boldsymbol{m}$ and $\boldsymbol{j}$, respectively characterizing input and output particle number configurations. Similarly as in Eq.~\eqref{eq:Muir2}, the convolution is between the two objects $B_{\boldsymbol{m} \to \boldsymbol{j}}$ and $(-1)^{|\boldsymbol{j}|}F_{\boldsymbol{m}\to\boldsymbol{j}}$. Notice again the presence of the sign $(-1)^{|\boldsymbol{j}|}$, which takes into account quantum interferences and depends on the total number of particles of each output configuration (and therefore also input configurations, as particle number is conserved, meaning that $\boldsymbol{m} = \boldsymbol{j}$ for $B_{\boldsymbol{m} \to \boldsymbol{j}}$ and $F_{\boldsymbol{m}\to\boldsymbol{j}}$ to be non-zero). This is to be compared with the classical convolution in Eq.~\eqref{eq: classical convolution}, which contains no such interference terms.
The convolution in Eq~.\eqref{eq: Complementarity BS/FS} can be understood to model the following process:
starting from a given input/output configuration $(\boldsymbol{k},\boldsymbol{i})$ for a given species (bosons or fermions), we then consider all the possible configurations $\{(\boldsymbol{m},\boldsymbol{j}):\boldsymbol{k}-\boldsymbol{m}\geq 0,\boldsymbol{i}-\boldsymbol{j}\geq 0\}$ in which some particles got transformed into the complementary species (bosons to fermions and fermions to bosons), with an added sign depending on total particle number and accounting for interferences.

\begin{remark}[Limiting cases] The case in which $S_{ij}=1$ for all $i,j$ (namely, for fully indistinguishable particles), reduces the complexity from a tensor permanent/determinant to their matrix versions.
In that case, Eq.~\eqref{eq:DetPer} simplifies to the following identity, which was recently proven in Ref.~\onlinecite{jabbour2023bosonfermion}:
\begin{equation}
    \sum_{\boldsymbol{m},\boldsymbol{j}\in\mathbb{N}^{n}}^{\boldsymbol{k},\boldsymbol{i}}(-1)^{|\boldsymbol{j}|}\frac{|\det(U^{(\boldsymbol{m},\boldsymbol{j})})|^{2}}{\boldsymbol{m}!\boldsymbol{j}!}\frac{|\perm(U^{(\boldsymbol{k-m},\boldsymbol{i-j})})|^{2}}{\boldsymbol{(k-m)}!\boldsymbol{(i-j)}!}
    =\delta_{\boldsymbol{k},\boldsymbol{0}}\delta_{\boldsymbol{i},\boldsymbol{0}}.
\end{equation}
In contrast, in the case where $S=I$ (\ie, for fully distinguishable particles), it can be shown that we indeed recover the usual convolution of Eq.~\eqref{eq: classical convolution}. We point the interested reader to Appendix~\ref{sec: classical fermions and bosons} for more details about such a case, and how the ``quantum identity'' indeed simplifies to the ``classical one''.
\end{remark}

\begin{figure}[b]
    \centering
    \includegraphics[width=0.8\linewidth]{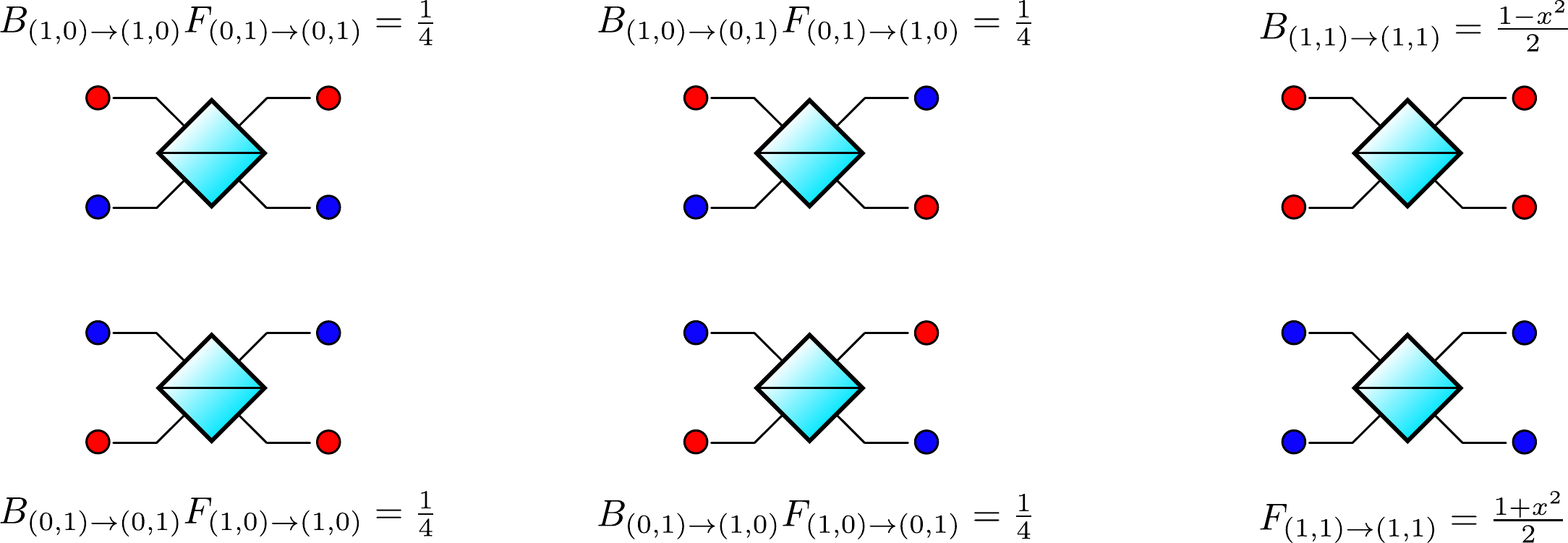}
    \caption{Schematics of the case of a 50:50 beam-splitter with single particle inputs (red for bosons and blue for fermions) given an overlap between the degrees of freedom $|\langle\phi_{1}|\phi_{2}\rangle|=x$. In this case, Theorem~\ref{cor: Boson/Fermion complementarity} can be written explicitly given the low number of possible combination. In particular, we would have: $\frac{1-x^{2}}{2} + \frac{1+x^{2}}{2} +(-1)\cdot4\cdot\frac{1}{4}=0$.}
    \label{fig: BS/FS}
\end{figure}

In Section~\ref{sec:interpret} below, we will discuss some applications of Theorem~\ref{cor: Boson/Fermion complementarity} in the context of quantum metrology. Before doing so, let us interpret Eq.~\eqref{eq: Complementarity BS/FS} further in physical terms, in particular the notion of complementarity between bosons and fermions it implies. To do so, we consider the simple case of two particles impinging on a two-mode interferometer, that is, a simple beam-splitter.
As already mentioned a few times, it is known that bosons have a tendency to bunch (\ie, they tend to occupy the same modes), while fermions tend to antibunch (\ie, they occupy different spacial modes), as a result of Pauli's exclusion principle. In the case of a 50:50 beam-splitter, such a behavior for bosons is quite significant in the history of quantum optics, since it gives rise to the celebrated Hong-Ou-Mandel effect~\cite{HOM}. 
We are going to show how Theorem~\ref{cor: Boson/Fermion complementarity} can be exploited to make this analysis more precise.
For that, we take the situation where one particle is impinging into each of the two arms of the beam-splitter, and we look at the probability of having a single particle at each of the output arms. This is in fact exactly the situation describing the Hong-Ou-Mandel effect.
In the case of a general beam-splitter (\ie, not necessarily 50:50), Eq.~\eqref{eq: Complementarity BS/FS} tell us that
\begin{equation}
    \begin{aligned}
        B_{(1,1)\to (1,1)}+F_{(1,1)\to (1,1)}&= B_{(0,1)\to(0,1)}F_{(1,0)\to(1,0)}+ B_{(1,0)\to(1,0)}F_{(0,1)\to(0,1)}\\
        & \qquad +B_{(0,1)\to(1,0)}F_{(1,0)\to(0,1)}+ B_{(1,0)\to(0,1)}F_{(0,1)\to(1,0)}.
    \end{aligned}
\end{equation}
Now, when a single particle is traveling through the beam-splitter, its nature is irrelevant as there is no interference involved, and we can replace the probabilities on the right-hand side with those we would get for classical particles. We thus obtain
\begin{equation}
    \begin{aligned}
        B_{(1,1)\to (1,1)}+F_{(1,1)\to (1,1)}&= P_{(0,1)\to(0,1)}P_{(1,0)\to(1,0)}+ P_{(1,0)\to(1,0)}P_{(0,1)\to(0,1)} \\
        & \qquad + P_{(0,1)\to(1,0)}P_{(1,0)\to(0,1)}+ P_{(1,0)\to(0,1)}P_{(0,1)\to(1,0)}\\
        &=2P_{(0,1)\to(0,1)}P_{(1,0)\to(1,0)}+ 2P_{(1,0)\to(0,1)}P_{(0,1)\to(1,0)}
    \end{aligned}
\end{equation}
so that we simply have
\begin{equation}\label{eq: complementarity 2x2}
    B_{(1,1)\to (1,1)}+F_{(1,1)\to (1,1)} = 2 P_{(1,1)\to (1,1)}.
\end{equation}
This simple, yet elegant relation tells us that the tendency of bosons to bunch in a mode (measured by $P_{(1,1)\to (1,1)} - B_{(1,1)\to (1,1)}$, \ie, the default of coincidence probability with respect to classical particles) is exactly the same as the tendency of fermions of antibunch in the same mode (measured by $F_{(1,1)\to (1,1)} - P_{(1,1)\to (1,1)}$)$)$.
The situation is illustrated on Fig.~\ref{fig: BS/FS} in the case of a 50:50 beam-splitter.

It is important to clarify that not all choices of a unitary $U$ give rise to the same bunching behavior. Let us consider a Haar random unitary of dimension $2$, that is an average over the set $U(2)$ of $2\times 2$ unitary matrices. We make use of the notion of Weingarten function~\cite{collins2022weingarten} in Appendix~\ref{sec:haar} to show the following:
\begin{align}
    \int_{U(2)}B_{(1,1)\to (1,1)} dU = \frac{2-|\langle \phi_{1}|\phi_{2}\rangle|^{2}}{3}, \quad \int_{U(2)}F_{(1,1)\to (1,1)} dU = \frac{2+|\langle \phi_{1}|\phi_{2}\rangle|^{2}}{3},\label{eq: Haar_prob1}\\
    \int_{U(2)}B_{(1,1)\to (2,0)} dU = \frac{1+|\langle \phi_{1}|\phi_{2}\rangle|^{2}}{6}, \quad \int_{U(2)}F_{(1,1)\to (2,0)} dU = \frac{1-|\langle \phi_{1}|\phi_{2}\rangle|^{2}}{6}.\label{eq: Haar_prob2}
\end{align}
As one can see, on average, bosons have a tendency to bunch as indistinguishability increases, while the opposite happens for fermions. What is interesting to notice, is that the sum of the bunching probabilities for bosons and fermions is constant, and does not depend on the overlap $|\langle \phi_{1}|\phi_{2}\rangle|^{2}$, which quantifies distinguishability and is the signature of two particle interference. In other words,
\begin{equation}\label{eq: Haar_prob3}
    \int_{U(2)} \left( B_{(1,1)\to (1,1)} +F_{(1,1)\to (1,1)} \right) dU = \frac{4}{3}.
\end{equation}
The same can be said about the antibunching probabilities, but with opposite behaviors for the two types of particles, namely,
\begin{equation}\label{eq: Haar_prob4}
    \int_{U(2)} \left( B_{(1,1)\to (2,0)} + F_{(1,1)\to (2,0)} \right) dU = \frac{1}{3}.
\end{equation}
Equations~\eqref{eq: Haar_prob3} and~\eqref{eq: Haar_prob4} really capture the notion of complementarity between the transition probabilities of bosons and fermions.
This can in fact most notably be understood through Eq.~\eqref{eq: complementarity 2x2}. Indeed, while its left-hand side depends on the overlap $|\langle \phi_{1}|\phi_{2}\rangle|^{2}$, the right-hand side, which depends only on classical quantities, exhibits no interferences and thus \emph{cannot} depend on $|\langle \phi_{1}|\phi_{2}\rangle|^{2}$.

The case of three particles is considerably harder to deal with, due to the fact that the number of terms appearing in Eq.~\eqref{eq: Complementarity BS/FS} grows exponentially with the number of particles considered. Nevertheless, it is useful to at least discuss the role played by the sign changes which give rise to interference, as well as complementarity.
It can be shown that the tensor permanent and determinant can be rewritten as follows:
\begin{equation}\label{eq: tensor perm/det compact}
    \begin{aligned}
        & \Perm(W) = \sum_{\sigma \in \mathcal{S}_n}\perm(U\odot \overline{U}_{\sigma})\prod_{i=1}^{n}S_{i,\sigma_{i}},\\
        & \Det(W) = \sum_{\sigma \in \mathcal{S}_n}\operatorname{sgn}(\sigma)\perm(U\odot \overline{U}_{\sigma})\prod_{i=1}^{n}S_{i,\sigma_{i}},
    \end{aligned}
\end{equation}
where $\overline{U}_{\sigma}$ denotes the complex conjugate of $U$, whose columns have been permuted according to $\sigma$ (see Ref.~\onlinecite{Tichy_2015} for details).
Consider for instance the transition probabilities $B_{(1,1,1)\to (1,1,1)}$ and $F_{(1,1,1)\to (1,1,1)}$.
By examining the above equations, one understands that those probabilities will in general depend on terms such as $S_{12}S_{23}S_{31}$ and $S_{13}S_{32}S_{21}$, which are a signature of three-particle interference. On the other hand, for such probabilities, Eq.~\eqref{eq: Complementarity BS/FS} allows us to write an identity of the following form:
\begin{equation}\label{eq:diff3}
    B_{(1,1,1)\to (1,1,1)}-F_{(1,1,1)\to (1,1,1)}=\text{``terms with  two particles interference''}.
\end{equation}
Since we are considering an odd number of particles, there is a minus sign appearing in front of one of the terms of the left-hand side of Eq.~\eqref{eq:diff3}. In other words, we obtain an equation constraining the \emph{difference} between the boson and fermion transition probabilities this time. Because of this, one understands that terms such as $S_{12}S_{23}S_{31}$ and $S_{13}S_{32}S_{21}$ are in fact canceled in the difference in Eq.~\eqref{eq:diff3}, so that it actually does not depend on such three-particle interference terms.

Evidently, the generalization to more than three particles becomes increasingly involved. There is however a clear structure in Eq.~\eqref{eq: Complementarity BS/FS}, which constrains the sum ``$B+F$'' of the bosonic and fermionic probabilities for even particle numbers, whereas it constraints their difference ``$B-F$'' for odd particle numbers.

\section{Physical implications for quantum metrology\label{sec:interpret}}

Beyond the transition probabilities themselves, the fundamental dichotomy between bosonic and fermionic manifests physically in the variance and covariance of their output particle number distributions. Understanding these multiparticle fluctuations is paramount for quantum technologies, in particular in quantum metrology~\cite{Barbieri_2022}, where the single-mode variance provides fundamental limits on the precision of phase estimation protocols, and in validation techniques for scattering experiment, where the variance can be directly related to witnesses of indistinguishability~\cite{rodari2024semideviceindependentcharacterizationmultiphoton}. The correlation matrices for bosonic, fermionic, and classical particle number distributions are respectively defined as follows:
\begin{align}
    C_{a,b}^{B}(\boldsymbol{k}) &=\sum_{\boldsymbol{i}}i_{a}i_{b}B_{\boldsymbol{k}\to \boldsymbol{i}}-{\bigg (}\sum_{\boldsymbol{i}}i_{a}B_{\boldsymbol{k}\to \boldsymbol{i}}{\bigg )}{\bigg (}\sum_{\boldsymbol{i}}i_{b}B_{\boldsymbol{k}\to \boldsymbol{i}}{\bigg )},\label{eq:covB}\\
    C_{a,b}^{F}(\boldsymbol{k}) &=\sum_{\boldsymbol{i}}i_{a}i_{b}F_{\boldsymbol{k}\to \boldsymbol{i}}-{\bigg (}\sum_{\boldsymbol{i}}i_{a}F_{\boldsymbol{k}\to \boldsymbol{i}}{\bigg )}{\bigg (}\sum_{\boldsymbol{i}}i_{b}F_{\boldsymbol{k}\to \boldsymbol{i}}{\bigg )},\label{eq:covF}\\
    C_{a,b}^{cl}(\boldsymbol{k}) &=\sum_{\boldsymbol{i}}i_{a}i_{b}P_{\boldsymbol{k}\to \boldsymbol{i}}-{\bigg (}\sum_{\boldsymbol{i}}i_{a}P_{\boldsymbol{k}\to \boldsymbol{i}}{\bigg )}{\bigg (}\sum_{\boldsymbol{i}}i_{b}P_{\boldsymbol{k}\to \boldsymbol{i}}{\bigg )},\label{eq:covcl}
\end{align}
where $cl$ stands for the classical distribution.
We will consider $\boldsymbol{k}\in\{0,1\}^{m}$ in what follows, in order to have a meaningful comparison between the different particle natures.
We will show that the behavior of bosons and fermions is antithetical in terms of covariance matrix. Namely, we have the following Theorem (see Section~\ref{sec:proofTheo2} for a proof).
\begin{theo}\label{th: Covariances}
    For every unitary $U\in U(m)$ and Gram matrix $S$, we have that
    \begin{equation}\label{eq:th: Covariances}
        C^{B}(\boldsymbol{k})+C^{F}(\boldsymbol{k})=2C^{cl}(\boldsymbol{k}),
    \end{equation}
    for all $\boldsymbol{k}\in \{0,1\}^{m}$.
\end{theo}
Equation~\eqref{eq:th: Covariances} is reminiscent of Eq.~\eqref{eq: complementarity 2x2}, but in contrast concerns covariance matrices that characterize any total number of particles $|\boldsymbol{k}|$. It is in fact striking that such covariance matrices for different particle natures are constrained by such a simple relation. Remember that the right-hand side of Eq.~\eqref{eq:th: Covariances} is constant in terms of the indistinguishability parameter (it does not depends on it), meaning that the sum on the left-hand side does not depend on the the indistinguishability parameter either, even though it characterizes bosonic and fermionic quantities that each depend on this parameter.
If we compute the covariance matrices for bosons and fermions, we obtain~\cite{walschaersStatisticalBenchmarkBosonSampling2016} (see Section~\ref{sec: Moment generating} for a derivation method)
\begin{align}
    C_{ij}^{B}(\boldsymbol{k}) = \begin{cases}
        \sum_{l} \delta_{k_{l},1}|U_{il}|^2 - \sum_{l} |U_{il}|^4\delta_{k_{l},1} + \sum_{l\neq l' }|S_{l,l'}|^2 |U_{il}|^2 |U_{il'}|^2 \delta_{k_{l},1}\delta_{k_{l'},1}  &\text{if} \ i=j,\\
        -\sum_{l}|U_{l,i}U_{l,j}|^{2}\delta_{k_{l},1} +\sum_{l\neq l'}|S_{l,l'}|^{2}\delta_{k_{l},1}\delta_{k_{l'},1}U_{l,i}U_{l',j}\bar{U}_{l',i}\bar{U}_{l,j} & \text{if} \ i\neq j,\\
    \end{cases}\label{eq:cova1}\\
    C_{ij}^{F}(\boldsymbol{k}) = \begin{cases}
        \sum_{l} \delta_{k_{l},1}|U_{il}|^2 - \sum_{l} |U_{il}|^4\delta_{k_{l},1} - \sum_{l\neq l' }|S_{l,l'}|^2 |U_{il}|^2 |U_{il'}|^2 \delta_{k_{l},1}\delta_{k_{l'},1} & \text{if} \ i=j,\\
        -\sum_{l}|U_{l,i}U_{l,j}|^{2}\delta_{k_{l},1} -\sum_{l\neq l'}|S_{l,l'}|^{2}\delta_{k_{l},1}\delta_{k_{l'},1}U_{l,i}U_{l',j}\bar{U}_{l',i}\bar{U}_{l,j} & \text{if} \ i\neq j.\\
    \end{cases}\label{eq:cova2}
\end{align}
The above is true for all partial distinguishability regime, which is quite interesting in the context of quantum metrology. In fact, it is know that for various experimental schemes (such as the Mach-Zehnder interferometer), indistinguishability plays a fundamental role in our ability to estimate phases~\cite{leeTwoparticleIndistinguishabilityIdentification2019}.

\begin{figure}[b]
    \centering
    \includegraphics[width=0.6\linewidth]{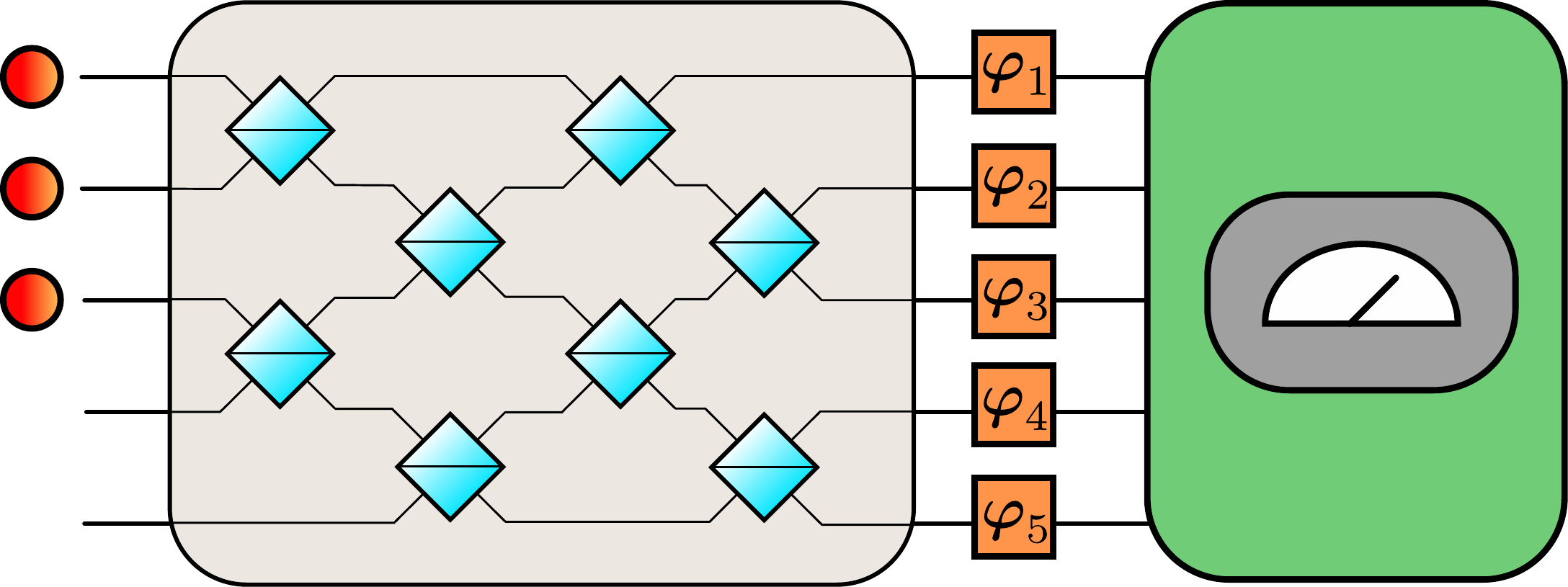}
    \caption{We consider a simple protocol where a set of particles (bosons, fermions, classical) interfere through a linear interferometer. The goal is to measure a quantity encoded by means of the phase shifter with phases $\varphi_{i}$. Any measurement can be performed on them.}
    \label{fig:QFI}
\end{figure}

Let us study some implications of Eq.~\eqref{eq:th: Covariances} in the paradigm of an experimentally relevant situation. We consider the general protocol shown on Fig.~\ref{fig:QFI}, where a set of particles are sent into a linear interferometer characterized by a matrix $U$. The goal is to measure some phases $\varphi_{i}$ that are encoded through phase shifters described by unitaries $e^{i\varphi_{i}\hat{n}_{i}}$, where $\hat{n}_{i}= \hat{a}_{i}^{\dagger} \hat{a}_{i}$.
The best estimator of $\varphi_{i}$ after $N$ measurements will have a variance $\Delta(\varphi_i)$ constrained by the Cramér-Rao bound, \ie,
 \begin{equation}\label{eq:CR}
     \Delta(\varphi_i)\geq \frac{1}{NF_{Q}(\varphi_{i})},
 \end{equation}
 where $F_{Q}(\varphi_{i})$ is the so-called quantum Fisher information~\cite{Barbieri_2022}.
 For a system in a pure state $|\Psi\rangle$, the quantum Fisher information can be computed as follows:
 \begin{equation}
    F_{Q}(\varphi)=4{\bigg (}\langle \partial_{\varphi}\Psi|\partial_{\varphi}\Psi\rangle -|\langle \partial_{\varphi}\Psi|\Psi\rangle|^{2}{\bigg )}.
\end{equation}
In the context of the protocol we consider here, it can be shown that
\begin{equation}\label{eq:FI}
    F_{Q}(\varphi_{i})=4 C_{ii}(\boldsymbol{k}),
\end{equation}
for each type of particle, \ie, bosonic, fermionic, and classical.
Now, we can make use of Theorem~\ref{th: Covariances} to compare the phase-estimation performances of bosonic and fermionic systems as a function of partial distinguishability. It was shown in Ref.~\onlinecite{rodari2024semideviceindependentcharacterizationmultiphoton} that, for bosonic systems, the variance is maximized by indistinguishable particles in an unbiased interferometer, namely when $|U_{ij}|=\frac{1}{\sqrt{m}}$.
It then follows from the combination of Eqs.~\eqref{eq:th: Covariances} and~\eqref{eq:FI} that indistinguishable fermions attain the minimum variance for the particle number distribution in the same setting. In other words, the regime in which bosons are optimal for phase estimation, with respect to partial distinguishability, corresponds to one where fermions exhibit a poor performance for phase estimation. This phenomenon was already observed in Ref.~\onlinecite{leeTwoparticleIndistinguishabilityIdentification2019}, where the relevant unitary was an unbiased beam splitter, but we see here that it carries out for any arbitrary interferometer.

\section{Proofs of the results\label{sec:proof}}

\subsection{Proof of Lemma~\ref{th: thermal GF}\label{sec:proofLemma}}

The method used to calculate $g_{B}$ and $g_{F}$ involves the use of the quantum characteristic function. Although the main idea behind this approach is the same for both cases, their formalism is intrinsically different. As a consequence, to prove Lemma~\ref{th: thermal GF}, we separate the bosonic case from the fermionic one.

\subsubsection{Bosonic Characteristic function\label{sec: Bosonic}}

We start by recalling the definition of quantum characteristic function of a state $\rho$, by first defining (for every vector of complex numbers $\boldsymbol{\alpha}=(\alpha_{1},...,\alpha_{m})^{T}$) the  symmetric displacement operator:
\begin{equation}
    D(\boldsymbol{\alpha}) = \exp( \boldsymbol{\hat{a}}^{\dagger}\boldsymbol{\alpha} -\boldsymbol{\alpha}^{\dagger}\boldsymbol{\hat{a}}) = \prod_{i=1}^m \exp(\alpha_{i} \hat{a}_{i}^{\dagger} -\alpha_{i}^{*} \hat{a}_{i}),
\end{equation}
where we used the notation $\boldsymbol{\hat{a}}=(\hat{a}_{1},...,\hat{a}_{m})^{T}$ with $\hat{a}_{i}$ the annihilation operator of the $i$\textsuperscript{th} mode. The symmetric characteristic function is defined as follows:
\begin{equation}
    \chi_{\rho}(\boldsymbol{\alpha}) = \tr{D(\boldsymbol{\alpha})\rho}.
\end{equation}
We will exploit two facts:
\begin{enumerate}
    \item The effect of a linear interferometer
    is described by Eqs.~\eqref{eq:inter} and~\eqref{eq:Uc},
    so that the evolution of an input state $\rho$ under the effect of the linear interferometer can be described by the following characteristic function:
\begin{equation}
    \chi_{\hat{U} \rho \hat{U}^{\dagger}}(\alpha)=\tr{D(\boldsymbol{\alpha}) \hat{U} \rho \hat{U}^{\dagger}}=\tr{D(U^{\dagger}\boldsymbol{\alpha})\rho}.
\end{equation}
\item The expectation value of the product of two operators $A$ and $B^{\dagger}$ can be directly computed from the characteristic function using the Parseval identity:
\begin{equation}
    \tr{AB^{\dagger}}=\int \frac{d^{2n}\alpha}{\pi^{n}}\chi_{A}(\boldsymbol{\alpha})\overline{\chi_{B}(\boldsymbol{\alpha})},
\end{equation}
where the bar over a complex number represents a complex-conjugate.
\end{enumerate}
We start from the case of indistinguishable particles, then later generalize the formalism in the presence of a Gram matrix $S$. The second property allows us to compute the transition probabilities. Let us start with an example and let us suppose $\rho$ is a single mode state. We are interested in the probability $\langle k|\rho|k\rangle$. This is equal to:
\begin{equation}
    \langle k|\rho|k\rangle=\tr{|k\rangle\langle k|\rho}=\int \frac{d^{2}\alpha}{\pi}\chi_{\rho}(\alpha)\chi_{|k\rangle\langle k|}(\alpha).
\end{equation}
To consider thermal states, we introduce the operator $\sigma(y)$ defined as:
\begin{equation}
    \sigma(y)=\sum_{n=0}^{\infty}y^{n}|n\rangle \langle n|.
\end{equation}
Thus,
\begin{equation}
    \langle k|\rho|k\rangle=\frac{1}{k!}\frac{\partial^{k}}{\partial y^{k}}\sum_{n=0}^{\infty}y^{n}\int \frac{d^{2}\alpha}{\pi}\chi_{\rho}(\alpha)\chi_{|n\rangle\langle n|}(\alpha){\bigg |}_{y=0}=\frac{1}{k!}\frac{\partial^{k}}{\partial y^{k}}\int \frac{d^{2}\alpha}{\pi}\chi_{\rho}(\alpha)\chi_{\sigma(y)}(\alpha){\bigg |}_{y=0}.
\end{equation}
Now, the characteristic function of a Fock state can be written as follows~\cite{Counting}:
\begin{equation}
    \chi_{|k\rangle\langle k|}(\alpha)=e^{-\frac{|\alpha|^{2}}{2}}L_{k}(|\alpha|^{2}),
\end{equation}
where $L_{k}$ indicates the $k-$th order Laguerre's polynomial. From this we can compute the characteristic function of $\sigma(y)$ as:
\begin{equation}
    \chi_{\sigma(y)}(\alpha)=e^{-\frac{|\alpha|^{2}}{2}}\sum_{n=0}^{\infty}y^{n}L_{n}(|\alpha|^{2})=e^{-\frac{|\alpha|^{2}}{2}}\frac{1}{1-y}e^{-\frac{y}{1-y}|\alpha|^{2}}.
\end{equation}
The straightforward generalization for a multimode state can be done by defining the operator
\begin{equation}
    \Omega(\boldsymbol{y}) = \bigotimes_{i=1}^m \sigma(y_{i}),
\end{equation}
for which we get
\begin{equation}
    \chi_{\Omega(\boldsymbol{y})}(\boldsymbol{\alpha})=e^{-\frac{|\boldsymbol{\alpha}|^{2}}{2}} \prod_{i=1}^m \frac{1}{1-y_{i}}e^{-\frac{y_{i}}{1-y_{i}}|\alpha_{i}|^{2}}.
\end{equation}
We can also derive in the same way the characteristic function of $\rho^{(b)}$, since we have that
\begin{equation}
    \rho^{(b)}(\boldsymbol{x})=\det(I-X)\Omega(\boldsymbol{x}) \quad \Rightarrow \quad \chi_{\rho^{(b)}(\boldsymbol{x})}=\det(I-X)\chi_{\Omega(\boldsymbol{x})}=e^{-\boldsymbol{\alpha}^{\dagger}(I/2 +X(I-X)^{-1})\boldsymbol{\alpha}}.
\end{equation}
We thus obtain
\begin{equation}
    g_{B}(\boldsymbol{x},\boldsymbol{y})=\tr{\hat{U}\rho^{(b)}(\boldsymbol{x})\hat{U}^{\dagger}\Omega(\boldsymbol{y})}=\int \frac{d^{2n}\alpha}{\pi^{m}}\chi_{\hat{U}\rho^{(b)}(\boldsymbol{x})\hat{U}^{\dagger}}(\boldsymbol{\alpha})\chi_{\Omega(\boldsymbol{y})}(\boldsymbol{\alpha}).
\end{equation}
We can then explicitly compute the integral as follows
\begin{align}
    \tr{\hat{U} \rho^{(b)}(\boldsymbol{x}) \hat{U}^{\dagger}\Omega(\boldsymbol{y})}
    & = \int \frac{d^{2m}\alpha}{\pi^{m}}\chi_{\rho(\boldsymbol{x})}(U^{\dagger}\boldsymbol{\alpha})\chi_{\Omega(\boldsymbol{y})}(\boldsymbol{\alpha})\\
    & = \frac{1}{\det(I-Y)} \int \frac{d^{2m}\alpha}{\pi^{m}}e^{-\boldsymbol{\alpha}^{\dagger}(\frac{I}{2}+UX(I-X)^{-1}U^{\dagger})\boldsymbol{\alpha}}e^{-\boldsymbol{\alpha}^{\dagger}(\frac{I}{2}+Y(I-Y)^{-1})\boldsymbol{\alpha}}\\
    & = \frac{1}{\det(I-Y)}\frac{1}{\det(I+UX(I-X)^{-1}U^{\dagger}+Y(I-Y)^{-1})}\\
    & = \frac{1}{\det(I-Y+UX(I-X)^{-1}U^{\dagger}(I-Y)+Y)}\\
    & = \frac{1}{\det(I+(I-Y)UX(I-X)^{-1}U^{\dagger})}\\
    & = \frac{\det(I-X)}{\det(I-X)\det(I+(I-Y)UX(I-X)^{-1}U^{\dagger})}\\
    & = \frac{\det(I-X)}{\det(I-X)\det(I+U^{\dagger}(I-Y)UX(I-X)^{-1})}\\
    & = \frac{\det(I-X)}{\det(I-X+U^{\dagger}(I-Y)UX)}\\
    & = \frac{\det(I-X)}{\det(I-X+X-U^{\dagger}YUX)}\\
    & = \frac{\det(I-X)}{\det(I-U^{\dagger}YUX)}
\end{align}

We can now turn to the general case of partially distinguishable particles, repeating the same calculations.
Note that one has to be careful about dimensions of the matrices involved, in particular the Gram matrix $S$.
Let us assume that $rank(S)=d\leq m$. Then there exists a decomposition of $S$ as $S=C^{\dagger}C: C\in \mathbb{C}^{m\times d}$. This decomposition is not unique. Nevertheless, we assume that we have performed a choice for $C$, keeping in mind that the choice will not affect the final result.
Following the ideas and conventions described around Eq.~\eqref{eq:partialdis}, we take into account the effect of partial distinguishability on each annihilation operator by writing it in terms of the matrix $C$ as follows:
\begin{equation}
a_{i}^{\dagger}=\sum_{k} c_{i,k}a_{i,k}^{\dagger},
\end{equation}
where $c_{i,k}$ is the component of $C$ on the $i$\textsuperscript{th} row and the $k$\textsuperscript{th} column.
The overall transformation effected by the interferometer can be then be written as~\cite{GSO}:
\begin{equation}\label{eq: tensor transform}
    a_{i,1}^{\dagger}\to \sum_{k,j}c_{i,k}U_{i,j}a_{j,k}^{\dagger}
\end{equation}
The above operation can be rewritten in terms of the Kharti-Rao tensor product $\circ$ (see Appendix~\ref{sec: Kharti-Rao} of its definition and properties) as
$\boldsymbol{\hat{a}}=\left(\hat{a}_{1,1}, \cdots, \hat{a}_{m,1}\right)^T \rightarrow M \boldsymbol{\hat{a}}$ for $M=U\circ C$.
Notice that $M\in \mathbb{C}^{md\times m}$, mapping $m$ creation/annihilation operators into $m \times d$ creation/annihilation operators.
We now repeat the same calculation for the generating function, but considering a $\boldsymbol{\gamma}=(\alpha_{1,1},\dots, \alpha_{m,d})$ in a larger space, and defining a new state $\Omega(\boldsymbol{y})$ as
\begin{equation}
    \Omega(\boldsymbol{y})=\bigotimes_{i=1}^{m}\sigma(y_{i}), \quad \sigma(y) = \bigotimes_{j=1}^{d}\sum_{n=0}^{\infty}y^{n}\lvert 0,\dots,\overset{\text{(j\textsuperscript{th})}}{n},\dots,0\rangle \langle 0,\dots,\overset{\text{(j\textsuperscript{th})}}{n},\dots,0|.
\end{equation}
In that case, the generating function can be written as 
\begin{equation}
    \chi_{\Omega(\boldsymbol{y})}(\boldsymbol{\gamma})=\frac{e^{-\frac{|\boldsymbol{\gamma}|^{2}}{2}}}{\det^{d}(I-Y)}e^{-\boldsymbol{\gamma}^{\dagger}(Y(I-Y)^{-1}\otimes \mathbb{1}_{d})\boldsymbol{\gamma}},
\end{equation}
where $\mathbb{1}_{d}$ denotes the identity matrix of dimension $d$, while
\begin{equation}
    \chi_{U\rho^{(b)}(\boldsymbol{x})U^{\dagger}}(\boldsymbol{\gamma})=e^{-\frac{|\boldsymbol{\gamma}|^{2}}{2}-\boldsymbol{\alpha}_{1}^{\dagger}(X(I-X)^{-1})\boldsymbol{\alpha}_{1}}.
\end{equation}
Proceeding as before,
\begin{align*}
    \tr{U\rho^{(b)}(\boldsymbol{x})U^{\dagger}\Omega(\boldsymbol{y})}
    & = \int \frac{d^{2}\boldsymbol{\gamma}}{\pi^{md}}\chi_{U\rho^{(b)}(\boldsymbol{x})U^{\dagger}}(\boldsymbol{\gamma})\chi_{\Omega(\boldsymbol{y})}(\boldsymbol{y})\\
    & = \int \frac{d^{2}\boldsymbol{\gamma}}{\pi^{md}} e^{-\frac{|\boldsymbol{\gamma}|^{2}}{2}-\boldsymbol{\gamma}^{\dagger}M(X(I-X)^{-1})M^{\dagger}\boldsymbol{\gamma}}\frac{e^{-\frac{|\boldsymbol{\gamma}|^{2}}{2}}}{\det^{d}(I-Y)}e^{-\boldsymbol{\gamma}^{\dagger}(Y(I-Y)^{-1}\otimes \mathbb{1}_{d})\boldsymbol{\gamma}}\\
    & = \frac{1}{\det^{d}(I-Y)}\frac{1}{\det(I\otimes \mathbb{1}_{d}+MX(I-X)^{-1}M^{\dagger} + (Y(I-Y)^{-1})\otimes \mathbb{1}_{d})}\\
    & = \frac{1}{\det^{d}(I-Y)}\frac{1}{\det(MX(I-X)^{-1}M^{\dagger} + (I-Y)^{-1}\otimes \mathbb{1}_{d})}\\
    & \stackrel{(a)}{=} \frac{1}{\det(I+MX(I-X)^{-1}M^{\dagger}[(I-Y)\otimes \mathbb{1}_{d}])}\\
    & = \frac{1}{\det(I+X(I-X)^{-1}M^{\dagger}[(I-Y)\otimes \mathbb{1}_{d}]M)}\\
    & = \frac{1}{\det(I+M^{\dagger}[(I-Y)\otimes \mathbb{1}_{d}]M X(I-X)^{-1})}\\
    & = \frac{1}{\det(I+[U^{\dagger}(I-Y)U\odot S] X(I-X)^{-1})}\\
    & \stackrel{(b)}{=} \frac{\det(I-X)}{\det(I-[(U^{\dagger}YU)\odot S]X)}
\end{align*}
In the above, $(a)$ follows from $(I+Y(I-Y)^{-1})=(I-Y)^{-1}$, while $(b)$ follows from identities on the Kharti-Rao product. In particular,
\begin{align}
    M^{\dagger}[(I-Y)\otimes \mathbb{1}_{d}]M
    & = (U^{\dagger}\bullet C^{\dagger})[(I-Y)\otimes \mathbb{1}_{d}](U\circ C)\\
    & = [U^{\dagger}(I-Y)\bullet (C^{\dagger}\mathbb{1}_{d})](U\circ C)\\
    & = (U^{\dagger}(I-Y)U)\odot S
\end{align}
and the rest of the derivation is similar to that of the indistinguishable case.

\subsubsection{Fermionic Characteristic function\label{sec: Fermionic dens}}

In the fermionic case, phase space variables, which we denote by $\beta_i$ (where $i$ is the mode index), are so-called anti-commuting Grassmann variables. They satisfy the following:
\begin{equation}
    \beta_{i}^{k}=0, \forall k \geq 2, \quad
    \{\beta_{i},\beta_{j}^{*}\}=0,\forall i,j = 1, \cdots, m.
\end{equation}
Following the same procedure as in the bosonic case, we start by introducing the single mode state $\rho(x)$ and $\sigma(y)$ defined as:
\begin{equation}
    \rho(x)=|0\rangle\langle 0|+x|1\rangle\langle 1|, \quad 
    \sigma(y)=|0\rangle\langle 0|+y|1\rangle\langle 1|.
\end{equation}
This time, due to Pauli's exclusion principle, we do not include Fock states with more than one particle.
From Parseval's relation, we get
\begin{equation}
    \tr{\rho(x)\sigma(y)}=\int d^{2}\beta e^{(1+yx)\beta\beta^{*}}=1+yx.
\end{equation}
The objects appearing in the integral are not characteristic functions, but the Parseval equation takes the form:
\begin{equation}
    \tr{\rho\sigma}=\int d^{2}\beta  K_{\rho}(\beta)W_{\sigma}(\beta),
\end{equation}
with
\begin{equation}
    K_{\rho}(\beta)=\frac{1}{2} + (1-2\langle \hat{n}\rangle_{\rho})\beta\beta^{*}, \quad
     W_{\sigma}(\beta)=\frac{1}{2} -\langle \hat{n}\rangle_{\sigma} + \beta\beta^{*},
\end{equation}
where $\hat{n}$ is the number operator (see Ref.~\onlinecite{DenFermi} for detailed derivations). To make things easier we can rewrite the product as:
\begin{equation}
\begin{aligned}
    \tr{\rho(x)\sigma(y)}=\int d^{2}\beta {\bigg (}\frac{1+x}{2}+\beta\beta^{*}(1-x){\bigg )}{\bigg (}\frac{1-y}{2}+\beta\beta^{*}(1+y){\bigg )}.
    \end{aligned}
\end{equation}
We can now use the properties of the Berezin integration and cancel all the terms which are not multiplied by $\beta\beta^{*}$, as those give a zero contribution to the integral, so that 
\begin{equation}
    \tr{\rho(x)\sigma(y)}=\int d^{2}\beta (1+yx)\beta\beta^{*}=\int d^{2}\beta e^{(1+yx)\beta\beta^{*}}.
\end{equation}
In the last step, we used properties of the Berezin integral for an exponential function. Since, for a linear interferometer, the relevant matrix transformation $U\in U(m)$ is independent of whether the particles are boson or fermions, the same calculations applied for the bosonic case can be directly applied here. In particular, we first introduce multimode states $\rho^{(f)}(\boldsymbol{x})$ and $\Omega(\boldsymbol{y})$ defined as follows:
\begin{align}
    \rho^{(f)}(\boldsymbol{x})&=\frac{1}{\det(I+X)}\bigotimes_{i=1}^{n}{\bigg (}|0\rangle\langle0|+x_{i}|1\rangle\langle1|{\bigg )},\\
    \Omega(\boldsymbol{y})&=\bigotimes_{i=1}^{n}{\bigg (}|0\rangle\langle0|+y_{i}|1\rangle\langle1|{\bigg )}.
\end{align}
Using those, we get:
\begin{equation}
    g_{F}(\boldsymbol{x},\boldsymbol{y})=\frac{1}{\det(I+X)}\int d^{2n}\boldsymbol{\beta} e^{\boldsymbol{\beta}^{\dagger}(\mathbb{1}+XY)\boldsymbol{\beta}}.
\end{equation}
We then use the change of variables $\beta_{i}\to \sum_{j}U_{i,j}^{\dagger}\beta_{j}$ to take into account the effect of the unitary transformation. This means that we can simply apply the change of variables $X\to UXU^{\dagger}$ to obtain:
\begin{equation}
    g_{F}(\boldsymbol{x},\boldsymbol{y})=\frac{\det(I+U^{\dagger}YUX)}{\det(I+X)}.
\end{equation}
Now, for the partially distinguishable particles case, repeating the same calculations as in the bosonic case, we finally obtain:
\begin{equation}
    g_{F}(\boldsymbol{x},\boldsymbol{y})=\frac{\det(I+(U^{\dagger}YU\odot S)X)}{\det(I+X)}.
\end{equation}

\subsection{Proof of Corollary~\ref{cor:Muir_new}\label{sec:proofCorr}}

Once we perform the change of variables $\boldsymbol{y}\to -\boldsymbol{y}$, we notice that 
 \begin{equation}
     g_{B}(\boldsymbol{x},-\boldsymbol{y})=\sum_{\boldsymbol{j}}b(\boldsymbol{x},\boldsymbol{j}) \, \boldsymbol{y}^{\boldsymbol{j}}(-1)^{|\boldsymbol{j}|}=\Bigg(\prod_{i=1}^m \frac{1-x_{i}}{1+x_{i}}\Bigg)\frac{1}{g_{F}(\boldsymbol{x},\boldsymbol{y})},
 \end{equation}
 and similarly,
  \begin{equation}
     g_{F}(\boldsymbol{x},-\boldsymbol{y})=\sum_{\boldsymbol{j}}f(\boldsymbol{x},\boldsymbol{j}) \, \boldsymbol{y}^{\boldsymbol{j}}(-1)^{|\boldsymbol{j}|}=\Bigg(\prod_{i=1}^m \frac{1-x_{i}}{1+x_{i}}\Bigg)\frac{1}{g_{B}(\boldsymbol{x},\boldsymbol{y})}.
 \end{equation}
We will make use of the convolution theorem, that states that given two discrete functions $\boldsymbol{p}$ and $\boldsymbol{q}$ with support in $\mathbb{N}$ and let $P(x)$ and $Q(x)$ their generating functions, we have that 
\begin{equation}
    P(x)Q(x)=\sum_{m}{\bigg (}\sum_{k}p_{k}\cdot q_{m-k}{\bigg )}x^{m},
\end{equation}
or, in other words, $P(x)Q(x)$ is the generating function of the convolution of $\boldsymbol{p}$ and $\boldsymbol{q}$. Given that the generating function of the distribution $\delta_{\boldsymbol{j},0}$ is $1$, we simply conclude that
\begin{equation}
     g_{F}(\boldsymbol{x},-\boldsymbol{y}) g_{B}(\boldsymbol{x},\boldsymbol{y})=\prod_{i=1}^m \frac{1-x_{i}}{1+x_{i}} \quad \Rightarrow \quad \sum_{\boldsymbol{i}}^{\boldsymbol{j}}(-1)^{|\boldsymbol{i}|}f(\boldsymbol{x},\boldsymbol{j}-\boldsymbol{i})b(\boldsymbol{x},\boldsymbol{i})=\Bigg(\prod_{i=1}^m \frac{1-x_{i}}{1+x_{i}}\Bigg)\delta_{\boldsymbol{j},0},
\end{equation}
while
\begin{equation}
     g_{F}(\boldsymbol{x},\boldsymbol{y}) g_{B}(\boldsymbol{x},-\boldsymbol{y})=\prod_{i=1}^m \frac{1-x_{i}}{1+x_{i}} \quad \Rightarrow \quad \sum_{\boldsymbol{i}}^{\boldsymbol{j}}(-1)^{|\boldsymbol{i}|}b(\boldsymbol{x},\boldsymbol{j}-\boldsymbol{i})f(\boldsymbol{x},\boldsymbol{i})=\Bigg(\prod_{i=1}^m \frac{1-x_{i}}{1+x_{i}}\Bigg)\delta_{\boldsymbol{j},0},
\end{equation}
which concludes the proof.

\subsection{Proof of Theorem~\ref{cor: Boson/Fermion complementarity}\label{sec:proofTheo}}

The proof follows the same idea as in the previous corollary.
The bosonic generating function can be written as 
\begin{equation}
    G_{B}(\boldsymbol{x},\boldsymbol{y})=\tr{U\Omega(\boldsymbol{x})U^{\dagger}\Omega(\boldsymbol{y})}=\frac{1}{\det(I-X)}\tr{U\rho^{(b)}(\boldsymbol{x})U^{\dagger}\Omega(\boldsymbol{y})}=\frac{1}{\det(I-X)}g_{b}(\boldsymbol{x},\boldsymbol{y}),
\end{equation}
where now we take two derivatives one with respect to the variables $\boldsymbol{x}$ (associated to the input) and one with respect to the variables $\boldsymbol{y}$ (for the output). With the same method we can retrieve the GF for fermions, namely,
\begin{equation}
    G_{B}(\boldsymbol{x},\boldsymbol{y})=\frac{1}{\det(I-(U^{\dagger}YU\odot S)X)}, \quad G_{F}(\boldsymbol{x},\boldsymbol{y})=\det(I+(U^{\dagger}YU\odot S)X).
\end{equation}
We can at this point observe that
\begin{equation}
    G_{B}(\boldsymbol{x},\boldsymbol{y})G_{F}(\boldsymbol{x},-\boldsymbol{y})=G_{B}(\boldsymbol{x},-\boldsymbol{y})G_{F}(\boldsymbol{x},\boldsymbol{y})=1.
\end{equation}
Notice that 
\begin{align}
    G_{B}(\boldsymbol{x},-\boldsymbol{y})=\sum_{\boldsymbol{m},\boldsymbol{j}}(-1)^{\boldsymbol{j}}B_{\boldsymbol{m}\to\boldsymbol{j}} \, \boldsymbol{x}^{\boldsymbol{m}}\boldsymbol{y}^{\boldsymbol{j}}, \quad
    G_{F}(\boldsymbol{x},-\boldsymbol{y})=\sum_{\boldsymbol{m},\boldsymbol{j}}(-1)^{\boldsymbol{j}}F_{\boldsymbol{m}\to\boldsymbol{j}} \, \boldsymbol{x}^{\boldsymbol{m}}\boldsymbol{y}^{\boldsymbol{j}}.
\end{align}
Following the same argument as before, we conclude the proof by making use of the convolution theorem.

\subsection{The moment generating function\label{sec: Moment generating}}

In this section, we introduce some details pertaining to the calculations of the covariances in Eqs.~\eqref{eq:cova1} and~\eqref{eq:cova2}.
To derive the covariance matrix, it is useful to first introduce the moment generating function, which for a random variable $X$ is defined as
\begin{equation}
    \mathcal{M}_{X}(t)=\mathbb{E}\left[e^{tX}\right].
\end{equation}
We recall that if $G_{X}(t)$ is the probability generating function of $X$, then $\mathcal{M}_{X}(t)=G_{X}(e^{t})$.
In the setting of Theorem~\ref{th: Covariances}, we are interested in fixed input configurations $\boldsymbol{k}\in \{0,1\}^{m}$. To derive the corresponding moment generating functions, it is thus useful to employ MacMahon's master theorem~\cite{MacMahon} and its counterpart for determinants (which is a particular case of the Cauchy-Binet formula~\cite{MarvinMinc}).
\begin{theo}\label{theo:MacMahon}
    Given $A\in \mathbb{C}^{m\times m}$ and $X=\diag(\boldsymbol{x})$,
    \begin{equation*}
        \perm(A^{\boldsymbol{j},\boldsymbol{j}})=[\boldsymbol{x}^{\boldsymbol{j}}] {\bigg (} \frac{1}{\det(I-XA)} {\bigg )}, \quad \det(A^{\boldsymbol{j},\boldsymbol{j}})=[\boldsymbol{x}^{\boldsymbol{j}}] {\bigg (} \det(I+XA) {\bigg )},
    \end{equation*}
    where $[\boldsymbol{x}^{\boldsymbol{j}}](f)$ represents the coefficient of the term $\prod_{i}x_{i}^{j_{i}}$ in the Taylor expansion of $f(\boldsymbol{x})$.
\end{theo}
Theorem~\ref{theo:MacMahon} implies the following for our GFs:
\begin{equation}
    G_{F}(\boldsymbol{y}|\boldsymbol{k}) = \sum_{\boldsymbol{i}}\boldsymbol{y}^{\boldsymbol{i}}F_{\boldsymbol{k}\to \boldsymbol{i}} = \det((U^{\dagger}YU\odot S)^{(\boldsymbol{k},\boldsymbol{k})}), \quad
    G_{B}(\boldsymbol{y}|\boldsymbol{k}) = \sum_{\boldsymbol{i}}\boldsymbol{y}^{\boldsymbol{i}}B_{\boldsymbol{k}\to \boldsymbol{i}} = \perm((U^{\dagger}YU\odot S)^{(\boldsymbol{k},\boldsymbol{k})}).
\end{equation}
A similar derivation can be found in Refs.~\onlinecite{anguita2025experimentalvalidationbosonsampling,Seron_2024}, where the generating functions were used as main tools in the context of the boson sampling validation task. For practical reasons, we rewrite the above bosonic generating function as follows:
\begin{equation}
    G_{B}(\boldsymbol{y}|\boldsymbol{k})=\perm((U^{\dagger}[I-I+Y]U\odot S)^{(\boldsymbol{k},\boldsymbol{k})})=\perm((I+U^{\dagger}[Y-I]U\odot S)^{(\boldsymbol{k},\boldsymbol{k})}),
\end{equation}
and similarly for fermions. By performing the substitution $Y\to e^{Y}$, we obtain the moment generating functions, which we denote as
\begin{equation}
    \mathcal{M}_{F}(\boldsymbol{y}|\boldsymbol{k})=G_{F}(e^{\boldsymbol{y}}|\boldsymbol{k}), \quad \mathcal{M}_{B}(\boldsymbol{y}|\boldsymbol{k})=G_{B}(e^{\boldsymbol{y}}|\boldsymbol{k}).
\end{equation}
As a consequence, we have that
\begin{equation}
    \mathcal{M}_{F}(\boldsymbol{y}|\boldsymbol{k}) = \det((I+U^{\dagger}[e^{Y}-I]U\odot S)^{(\boldsymbol{k},\boldsymbol{k})}), \quad \mathcal{M}_{B}(\boldsymbol{y}|\boldsymbol{k}) = \perm((I+U^{\dagger}[e^{Y}-I]U\odot S)^{(\boldsymbol{k},\boldsymbol{k})}).
\end{equation}
In particular, the moment generating functions satisfy
\begin{equation}
    \langle \prod_{i}\hat{n}_{i}^{j_{i}}\rangle=\frac{\partial^{\boldsymbol{j}}}{\partial \boldsymbol{y}^{\boldsymbol{j}}}\mathcal{M}_{B/F}(\boldsymbol{y}|\boldsymbol{k}){\bigg |}_{\boldsymbol{y}=0}.
\end{equation}

\subsection{Proof of Theorem~\ref{th: Covariances}\label{sec:proofTheo2}}

The bosonic covariance matrix of Eq.~\eqref{eq:covB} can be computed starting from the moment generating function as follows:
\begin{equation}
    C_{ij}^{B}=\frac{\partial^{2} }{\partial y_{i}\partial y_{j}}\mathcal{M}_{B}(\boldsymbol{y}|\boldsymbol{k}){\bigg |}_{\boldsymbol{y}=0} - \left( \frac{\partial }{\partial y_{i}}\mathcal{M}_{B}(\boldsymbol{y}|\boldsymbol{k}){\bigg |}_{\boldsymbol{y}=0}\right)\left( \frac{\partial }{\partial y_{j}}\mathcal{M}_{B}(\boldsymbol{y}|\boldsymbol{k}){\bigg |}_{\boldsymbol{y}=0}\right).
\end{equation}
The calculation is of course similar for the fermionic covariance matrix of Eq.~\eqref{eq:covF}.
We start by noticing that, if we define the object $\Gamma=(U^{\dagger}[e^{Y}-I]U\odot S)^{(\boldsymbol{k},\boldsymbol{k})}$, then 
\begin{equation}
        \Gamma_{ij}=S_{ij}\sum_{r: k_{r}=1}(e^{y_r}-1)U_{r i}^{*} U_{r j}.
\end{equation}  
We make use of the following property~\cite{minc1984permanents,grinberg2022principal} for any matrix $\Gamma\in \mathbb{C}^{m\times m}$:
\begin{equation}\label{eq: perm/det expansion}
    \det(I+\Gamma)=\sum_{\boldsymbol{j}\in \{0,1\}^{m}}\det(\Gamma^{(\boldsymbol{j},\boldsymbol{j})}), \quad
    \perm(I+\Gamma)=\sum_{\boldsymbol{j}\in \{0,1\}^{m}}\perm(\Gamma^{(\boldsymbol{j},\boldsymbol{j})}).
\end{equation}
The function $f(y)=e^{y}-1$ satisfies
\begin{equation}
    \frac{d^{k}}{dy^{k}}(e^{y}-1)^{l}=l!S(k,l),
\end{equation}
where the $S(k,l)$ are the so-called Stirling numbers of the second kind~\cite{OEISA008277}, which satisfy $S(k,l)=0$ if and only if $l>k$. This means that if we are interested in the second derivatives of the functions $\perm(I+\Gamma)$ or $\det(I+\Gamma)$ in the calculation of second moments, we can limit ourselves to terms of order up to $2$ in the corresponding series. We thus write the following:
\begin{equation}\label{eq1}
    \perm(I+\Gamma)+\det(I+\Gamma)=2+2\tr{\Gamma}+2\sum_{i>j}\Gamma_{ii}\Gamma_{jj}+\dots,
\end{equation}
where we made use of Eq.~\eqref{eq: perm/det expansion}, collecting all terms such that $|\boldsymbol{j}|\leq 2$. Notice that the cross terms ($\Gamma_{ij}\Gamma_{ji}$) cancel out since $\det$ and $\perm$ of a $2\times 2$ matrix have opposite signs multiplying such terms. For the classical case we can choose to use either of the fermionic or bosonic formalisms, since when $S_{i,j}=\delta_{i,j}$, the matrix $\Gamma$ is diagonal, and the permanent and the determinant of a diagonal matrix are the same. If we define the classical matrix $\Gamma^{cl}=\diag(\{\Gamma_{ii}\})$ then the expansion reads
\begin{equation}\label{eq2}
    \det(I+\Gamma^{cl})=\perm(I+\Gamma^{cl})=1+\sum_{i}\Gamma_{ii}+\sum_{i>j}\Gamma_{ii}\Gamma_{jj}+\dots.
\end{equation}
We this have that Eqs.~\eqref{eq1} and~\eqref{eq2} are equal when the series are truncated to the second order.
We thus get that the sum of the two covariance matrices for bosons and fermions is the same as twice the covariance matrix of the classical case, which proves the result.

\section{Conclusion\label{sec:conc}}

We have established complementarity relations that govern the behavior of transition probabilities of bosons and fermions within arbitrary linear interferometers, taking into account partial distinguishability. The latter is relevant in current experiments in which the systems' quantum properties play a key role, in particular in situations in which a quantum advantage is being demonstrated for practical purposes.
An important example is the so-called boson sampling paradigm~\cite{bosonsampling}, which investigates the computational complexity of simulating the scattering of many identical bosons through a multimode linear interferometer, and in which quantum interference plays a crucial role.
To prove our results, we exploited the formalism of the generating functions for transition probabilities. We thus began by showing what can be understood as a generalization of the MacMahon theorem, this time incorporating the notion of partial distinguishability (Lemma~\ref{th: thermal GF}).
This in turn provided a novel proof of a result of Muir dating from the end of the 19th century, in the form of Corollary~\ref{cor:Muir_new}. Most notably, it also allowed us to make a clear connection between Muir's result and the experiments of thermal boson/fermion sampling.
We then moved to our main result, Theorem~\ref{cor: Boson/Fermion complementarity}, a fundamental relation expressing a duality between bosonic and fermionic particle statistics in an arbitrary linear interferometer and subject to partial distinguishability, thus generalizing a result of Ref.~\onlinecite{jabbour2023bosonfermion}.
This relation happens to stem from a heretofore unknown mathematical identity between the determinants and permanents of tensors of order 3, in the form of Equation~\eqref{eq:DetPer}.

Beyond these fundamental multilinear algebra identities, we showed that these complementarity relations impose constraints on particle-number variances, forcing a fundamental sum rule: for any linear unitary network, the sum of the bosonic and fermionic covariance matrices is exactly twice the covariance matrix for particles behaving in classical fashion (Theorem~\ref{th: Covariances}). This novel variance constraint directly limits the fundamental precision bounds in quantum metrology protocols, such as phase estimation protocols. The relation leads to consequences that can be interpreted as follows: whenever increasing particles indistinguishability in bosonic systems minimizes the uncertainty on phase estimation, the opposite happens for fermionic systems.

We conclude with some open questions. A first relevant question is whether a similar notion of boson--fermion complementarity is exhibited when particles travel in an active medium, that is, one that does not conserve particle number, such as a parametric amplifier. Relatively recently, identities that can now be understood as originating from boson--fermion complementarity relations were proven in Ref.~\onlinecite{Jabbour2021} for such active transformations. It is thus natural to ask whether our present results can be generalized for active transformation, and what it implies in terms of matrix permanents and determinants.
Another question relates to the so-called quantum central limit theorem (QCLT), which concerns the equilibrium properties of subsystems of large bosonic systems. The implications of partial distinguishability on the QCLT were recently investigated in Ref.~\onlinecite{Robbio2024}. It would be interesting to explore the implications of our results in the context of the QCLT, in particular what they imply for the equilibrium properties of large fermionic systems.
A third question concerns suppression laws, or zero transmission laws, in linear interferometers, first introduced in Ref.~\onlinecite{tichy2010zero}, namely the existence of forbidden input-output transitions in such interferometers with certain symmetries. In a recent work (Ref.~\onlinecite{novo2026nativelinearopticalprotocolefficient}), suppression laws in bosonic systems were shown to be related to the estimation of the elements of the Gram matrix. It is therefore natural to look for the corresponding behavior in fermionic systems, and in particular to understand how bosonic and fermionic suppression laws are related as a consequence of their complementarity relations.
Finally, one may study the consequence of our relations on the complexity of simulating linear optical circuits. The boson sampling paradigm  derives from the computational hardness of computing permanents. Thus, the connection with determinants we obtain from Theorem~\ref{cor: Boson/Fermion complementarity} may have intriguing ramifications.

\begin{acknowledgments}
M.R. is a FRIA grantee of the Fonds de la Recherche Scientifique – FNRS (Belgium).
M.G.J. acknowledges funding from l’Agence Nationale de la Recherche (ANR, France) under project ANR-25‑CE47‑4015.
N.J.C. acknowledges support by the Fonds de la Recherche Scientifique – FNRS (Belgium) under project CHEQS within the Excellence of Science (EOS) program.
\end{acknowledgments}

\section*{Data Availability Statement}

Data sharing is not applicable to this article as no new data were created or analyzed in this study.

\appendix

\section{Khatri–Rao product and its properties\label{sec: Kharti-Rao}}

We introduce the following operations:
\begin{itemize}
    \item $A\circ B$ as the column-wise Kronecker product;
    \item $A\bullet B$ as the row-wise Kronecker product.
\end{itemize}
These are two particular cases of the Khatri-Rao product (also known as Block wise Kronecker product). Let $A$ and $B$ be two complex matrices and let us define a block partition for the rows and columns as follows:
\begin{equation}
    A=\begin{pmatrix}
        A_{1} & A_{2} & A_{3}
    \end{pmatrix}=\begin{pmatrix}
        a_{1} \\ a_{2} \\ a_{3}
    \end{pmatrix}, \quad 
    B=\begin{pmatrix}
        B_{1} & B_{2} & B_{3}
    \end{pmatrix}=\begin{pmatrix}
        b_{1} \\ b_{2} \\ b_{3}
    \end{pmatrix}.
\end{equation}
We used capital letters to indicate the column partition and lower cases for the row partition. Then, the row- and column-wise Kronecker products are respectively defined as:
\begin{equation}
    A\circ B=\begin{pmatrix}
        A_{1}\otimes B_{1} & A_{2}\otimes B_{2} & A_{2}\otimes B_{2}
    \end{pmatrix}, \quad
    A\bullet B=\begin{pmatrix}
        a_{1}\otimes b_{1} \\ a_{2}\otimes b_{2} \\ a_{2}\otimes b_{2}
    \end{pmatrix}.
\end{equation}
In our derivations, we make use of the following properties of the the above operations (for more details, see, Ref.~\onlinecite{face}):
\begin{enumerate}
    \item $(A\circ B)^{T}=A^{T}\bullet B^{T}$;
    \item $(A\bullet B)(A'\circ B)=(AA')\odot (BB')$;
    \item $(A\otimes B)(A'\circ B')=(AA')\circ(BB')$;
    \item $(A\bullet B)(A'\otimes B')=(AA')\bullet (BB')$.
\end{enumerate}
These operations find applications in various fields, such as artificial intelligence and machine learning, as they can be used to simplify calculations of convolutions.

\section{Classical particles\label{sec: classical fermions and bosons}}

In this section, we discuss the case of fully distinguishable particles in more details. In particular, we study why fully distinguishable bosons and fermions behave identically (up to the constrain due to the Pauli exclusion principle), and why we recover permanents in both expressions~\eqref{eq: BS transitions} and~\eqref{eq: FS transitions} even though we start from a determinant in the fermionic case.
We do so by working with the generating functions defined in Eqs.~\eqref{eq: thermal Gf} and ~\eqref{eq: thermal Gb}, which we rewrite here:
\begin{align}
    g_{F}(\boldsymbol{x},\boldsymbol{y}) &=\frac{\det(I+(U^{\dagger}YU\odot S)X)}{\det(I+X)},\label{eq: thermal Gf2}\\
    g_{B}(\boldsymbol{x},\boldsymbol{y}) &=\frac{\det(I-X)}{\det(I-(U^{\dagger}YU\odot S)X)}.\label{eq: thermal Gb2}
\end{align}
For distinguishable particles, transition probabilities are characterized by a Gram matrix $S=I$, for which we get
\begin{equation}
    (U^{\dagger}YU\odot I)_{i,j}=\delta_{i,j}\sum_{k}U_{k,i}^{*}U_{k,j}y_{k},
\end{equation}
which implies that the matrix $A=(U^{\dagger}YU\odot I)X$ appearing in the determinant in the numerator of~\eqref{eq: thermal Gf2} and the denominator of~\eqref{eq: thermal Gb2} can be written as a diagonal matrix whose diagonal elements satisfy:
\begin{equation}
    A_{i,i}=x_{i}\sum_{k}y_{k}|U_{i,k}|^{2}.
\end{equation}
Let us first consider the fermionic case, for which
\begin{equation}
    \det(I+A)=\prod_{i}(1+A_{i,i}).
\end{equation}
As we can see, the generating function $g_{F}(\boldsymbol{x},\boldsymbol{y})$ can be understood to characterize the convolution of independent variables, meaning that particles do not exhibit any quantum interference.
We can also understand the presence of the Pauli exclusion principle through the fact that any initial state with mode occupation vector $\boldsymbol{m}$, such that $m_{i}\geq 2$, will lead to a null transition probability, since each monomial appearing is of order one in the variables $x_{i}$.
The same can be said for the bosonic case, but this time we have:
\begin{equation}
    \frac{1}{\det(I-A)}=\frac{1}{\prod_{i} (1-A_{i,i}) }.
\end{equation}
This time the Pauli exclusion principle does not play a role since we have rational functions. Nevertheless, the generating function $g_{B}(\boldsymbol{x},\boldsymbol{y})$ can still be understood to describe the convolution of independent probability distributions.

It is interesting to notice that :
\begin{equation}
    \frac{\partial}{\partial x}\frac{1}{1-x}{\bigg |}_{x=0}=\frac{\partial}{\partial x}(1+x){\bigg |}_{x=0} \quad \Rightarrow \quad \frac{\partial^{\boldsymbol{j}}}{\partial \boldsymbol{x}^{\boldsymbol{j}}}\frac{1}{\det(I-A)}{\bigg |}_{\boldsymbol{x}=0}=\frac{\partial^{\boldsymbol{j}}}{\partial \boldsymbol{x}^{\boldsymbol{j}}}\det(I+A){\bigg |}_{\boldsymbol{x}=0}, \quad \forall \boldsymbol{j}\in \{0,1\}^{m}.
\end{equation}
This can be translated into the fact that, in the case of fully distinguishable particles, transition probabilities for fermionic and bosonic systems are the same (if the input occupation modes vector is feasible for fermions). This explains why, for $S=I$, both distributions $B_{\boldsymbol{i}\to\boldsymbol{k}}$ and $F_{\boldsymbol{i}\to\boldsymbol{k}}$ converge to the same distribution $P_{\boldsymbol{i}\to\boldsymbol{k}}$ written in terms of permanents.

As a consequence of Theorem~\ref{cor: Boson/Fermion complementarity}, we see that classical particles give rise to a complementarity relation
\begin{equation}
\sum_{\boldsymbol{m},\boldsymbol{j}\in\mathbb{N}^{n}}^{\boldsymbol{k},\boldsymbol{i}}(-1)^{|\boldsymbol{j}|}\frac{\perm(|U^{(\boldsymbol{m},\boldsymbol{j})}|^{2})}{\boldsymbol{m}!\boldsymbol{j}!}\frac{\perm(|U^{(\boldsymbol{k-m},\boldsymbol{i-j})}|^{2})}{\boldsymbol{(k-m)}!\boldsymbol{(i-j)}!}=\delta_{\boldsymbol{k},\boldsymbol{0}}\delta_{\boldsymbol{i},\boldsymbol{0}}.
\end{equation}
In fact, since only permanents appear in the above expression, it is easy to prove it by exploiting the fact that classical particles do not exhibit quantum interference.
Let us rewrite the identity in terms of probabilities, namely:
\begin{equation}\label{eq classic form}
    \sum_{\boldsymbol{m},\boldsymbol{j}\in\mathbb{N}^{n}}^{\boldsymbol{k},\boldsymbol{i}}(-1)^{|\boldsymbol{j}|}P_{\boldsymbol{m}\to\boldsymbol{j}}P_{\boldsymbol{k}-\boldsymbol{m}\to\boldsymbol{i}-\boldsymbol{j}}=\delta_{\boldsymbol{k},\boldsymbol{0}}\delta_{\boldsymbol{i},\boldsymbol{0}}.
\end{equation}
We wish to show the above relation in a simple way. We begin by making use of the fact that classical particles are characterized by convolutions of the following form:
\begin{equation}\label{Classical probabilities relation}
    P_{\boldsymbol{k}\to \boldsymbol{i}}=\sum_{\boldsymbol{m}}^{\boldsymbol{k}}P_{\boldsymbol{m}\to \boldsymbol{j}}P_{\boldsymbol{k-m}\to \boldsymbol{i-j}}, \quad \forall \boldsymbol{j}\leq \boldsymbol{i}.
\end{equation}
Simply summing terms of the above form, each multiplied by the suitable sign $(-1)^{|\boldsymbol{j}|}$, we get
\begin{equation}
    \sum_{\boldsymbol{j}\in\mathbb{N}^{n}}^{\boldsymbol{i}}(-1)^{|\boldsymbol{j}|}P_{\boldsymbol{k}\to \boldsymbol{i}}=P_{\boldsymbol{k}\to \boldsymbol{i}}\sum_{\boldsymbol{j}\in\mathbb{N}^{n}}^{\boldsymbol{i}}(-1)^{|\boldsymbol{j}|}=\delta_{\boldsymbol{k},0}\delta_{\boldsymbol{i},0}.
\end{equation}
Notice that $\sum_{j\in\mathbb{N}^{n}}^{i}(-1)^{|\boldsymbol{j}|}=\delta_{\boldsymbol{i},0}$ and thus $P_{\boldsymbol{k}\to \boldsymbol{i}}$ is not zero only for $\boldsymbol{k}=0$, which ends the proof.

\section{Computation of the Haar random average\label{sec:haar}}

The purpose of this section is to give a proof of Eqs.~\eqref{eq: Haar_prob1} and~\eqref{eq: Haar_prob2}. To do so, we start from the definitions of the transition probabilities for bosons and fermions
\begin{equation}
    \begin{aligned}\label{eq: (1,1) prob}
        B_{(1,1)\to (1,1)}&=|U_{1,1}U_{2,2}|^{2}+|U_{1,2}U_{2,1}|^{2} + 2|\langle \phi_{1}|\phi_{2}\rangle|^{2}\Re\left\{U_{1,1}U_{2,2}\overline{U_{1,2}U_{2,1}}\right\},\\
        F_{(1,1)\to (1,1)}&=|U_{1,1}U_{2,2}|^{2}+|U_{1,2}U_{2,1}|^{2} - 2|\langle \phi_{1}|\phi_{2}\rangle|^{2}\Re\left\{U_{1,1}U_{2,2}\overline{U_{1,2}U_{2,1}}\right\}.
    \end{aligned}
\end{equation}
We then recall that we can compute averages over the set of unitary matrices using the following identity~\cite{collins2022weingarten}
\begin{equation}
    \int_{U(m)} \prod_{r=1}^k U_{i_r j_r} \overline{U_{i'_r j'_r}} \, dU=\sum_{\sigma, \tau \in S_k}\prod_{r=1}^k \delta_{i_r, i'_{\sigma(r)}} \,\delta_{j_r, j'_{\tau(r)}}\,\mathrm{Wg}(\sigma^{-1}\tau,m),
\end{equation}
where $\mathrm{Wg}(\sigma^{-1}\tau,m)$ is the so-called Weingarten function. We thus have
\begin{equation}
    \begin{aligned}
        \int_{U(m)} dU U_{ij} U_{k\ell} \bar U_{mn} \bar U_{pq}
        &= (\delta_{im}\delta_{jn} \delta_{kp}\delta_{\ell q} + \delta_{ip}\delta_{jq} \delta_{km}\delta_{\ell n} ) \mathrm{Wg}(e,m)+ (\delta_{im} \delta_{jq} \delta_{kp}\delta_{\ell n} + \delta_{ip} \delta_{jn} \delta_{km}\delta_{\ell q}) \mathrm{Wg}((12),m),
    \end{aligned}
\end{equation}
where $e$ stands for the identity permutation while $(12)$ is a common notation for the transposition of elements $1$ and $2$. We recall that
\begin{align}
     \mathrm{Wg}(e,m) =\frac{1}{m^{2}-1}, \quad
      \mathrm{Wg}((12),m) =-\frac{1}{m(m^{2}-1)}.
\end{align}
Using the above in Eq.~\eqref{eq: (1,1) prob} gives us the identities in Eqs.~\eqref{eq: Haar_prob1} and~\eqref{eq: Haar_prob2}.

\bigskip

\bibliography{asa}

\end{document}